\documentclass[12pt]{article}

\usepackage[T1]{fontenc}
\usepackage{tipa}
\usepackage{mathtools} 
\usepackage{semantic}
\usepackage{cancel}
\usepackage{cite}
\mathlig{->}{\rightarrow}
\mathlig{|->}{\mapsto}
\usepackage{breqn}
\usepackage{graphicx}
\usepackage{epstopdf}
\usepackage{amsmath}
\usepackage{amssymb}
\usepackage{amsfonts}
\usepackage{amsthm}
\usepackage[usenames]{color}
\usepackage{array}
\usepackage[percent]{overpic}

\usepackage[
colorlinks={},
linkcolor=blue,
urlcolor=blue,
filecolor=blue,
citecolor=blue,
pdfstartview=FitV,
pdftitle={},
pdfauthor={},
pdfsubject={},
pdfkeywords={},
pdfpagemode=None,
bookmarksopen=true
,linktoc=page]{hyperref}

\usepackage{float}
\usepackage[utf8]{inputenc}
\usepackage{epsfig}
\usepackage{textcomp}
\usepackage{setspace}

\usepackage{comment}
\usepackage{amsmath,amssymb,extarrows,mathtools,graphicx,subfigure,setspace}

\usepackage{slashed}
\usepackage{color}
\usepackage{tikz}
\usepackage{fancyhdr}
\usetikzlibrary{decorations.pathmorphing}
\makeatother

\newcommand{\bea}{\begin{eqnarray}}
\newcommand{\eea}{\end{eqnarray}}
\newcommand{\ba}{\begin{array}}
	\newcommand{\ea}{\end{array}}
\newcommand{\ee}{\end{equation}}

\numberwithin{equation}{section}

\begin{document}
\begin{flushright}
	\texttt{}
\end{flushright}

\begin{centering}
\thispagestyle{empty}

	\textbf{\Large{
On JT gravity path integrals and the   tunneling process  }}
	
	\vspace{1.4cm}
	
	{\large Hamed Zolfi$^{a,b}$  and  Mohsen Alishahiha$^{c}$ }
	
	\vspace{0.9cm}
	
	\begin{minipage}{.9\textwidth}\small
		\begin{center}
			
	{
	$^{a}$~Department of Physics, College of Science, Shiraz University, Shiraz 71454, Iran	\\
 $^{b}$~School of Particles and Accelerators,
					Institute for Research in Fundamental Sciences (IPM)
					P.O. Box 19395-5531, Tehran, Iran
                    \\$^{c}$~School of Quantum Physics and Matter, Institute for Research in Fundamental Sciences (IPM),
		P.O. Box 19395-5531, Tehran, Iran
		}	
			
		\vspace{0.9cm}
			{\tt  \ Emails: h.zolfi@saadi.shirazu.ac.ir, alishah@ipm.ir }
		\\ 
			
		\end{center}
	\end{minipage}
	\vspace{1.5cm}

	\begin{abstract}
The Jackiw-Teitelboim (JT) gravity path integral of the trumpet can be interpreted as a transition amplitude from an older black hole to a younger one, accompanied by the emission of a baby universe, represented by the geodesic boundary of the trumpet. However, this interpretation becomes less straightforward for geometries with higher genus and multiple geodesic boundaries. In this paper, we examine the path integral for these more complex geometries and find that maintaining this interpretation requires accounting for a portion of the moduli space.

	\end{abstract}
\end{centering}
\newpage
\doublespacing
\tableofcontents
\setstretch{1.1}
\setcounter{equation}{0}
\setcounter{page}{2}

\section{Introduction}\label{intro}

The role of spacetime wormholes in the study of quantum gravity has attracted considerable attention in recent years, particularly about the Mathur/AMPS firewall paradox, which raises fundamental questions about the black hole information paradox and the nature of spacetime
\cite{haw, Lavrelashvili:1987jg, Maldacena:2004rf, Arkani-Hamed:2007cpn, Saad:2018bqo,Mathur:2009hf, Almheiri:2012rt, Bousso:2012as, Nomura:2012sw, Verlinde:2012cy, Papadodimas:2012aq}.

Within the framework of JT gravity \cite{JACKIW1985343, Teitelboim:1983ux, Nojiri:2024ycf, yang, kit, Bagrets:2017pwq, har}\footnote{JT gravity is not dual to an ordinary quantum system on the asymptotic boundary of spacetime, as one might expect from prior holographic duality, but rather to a matrix model, which is a random ensemble of quantum mechanical systems \cite{Saad:2019lba, Stanford:2019vob, Witten:2020wvy}.}, intriguing connections have emerged between wormholes and firewalls. These connections suggest that the formation of wormholes could lead to the generation of firewalls through the emission of large baby universes at late times \cite{Stanford:2022fdt, Saad:2019pqd}\footnote{It is worth noting that Coleman, Giddings, and Strominger established the connection between the physics of spacetime wormholes and baby universes in the 1980s \cite{Giddings:1987cg, COLEMAN1988867}.}. The emission of a baby universe from the wormhole causes a shortening of its length, which in turn leads to the formation of a firewall. Black hole firewalls exhibit characteristics similar to those of white holes \cite{Susskind:2015toa}, suggesting a compelling scenario in which an aging black hole could tunnel into a white hole or firewall by emitting large baby universes. The exploration of this phenomenon has been conducted within the context of JT gravity, initially for genus-one topologies, where the emission is associated with the production of a single baby universe \cite{Stanford:2022fdt}. Subsequent investigations have extended this analysis to higher-genus scenarios, revealing that the emission can involve an arbitrary number of baby universes \cite{Zolfi:2024ldx}.

Since the volume of a black hole's interior (or wormhole) is linearly proportional to its age \cite{Hartman:2013qma, Susskind:2014rva, Iliesiu:2021ari, Alishahiha:2022kzc, Alishahiha:2022exn, Iliesiu:2024cnh, Belin:2021bga, Blommaert:2024ftn, Bhattacharyya:2025gvd}\footnote{The time dependence of a 2+1 dimensional Lorentzian wormhole with three AdS boundaries \cite{Skenderis:2009ju} demonstrates non-linear growth that eventually saturates at late times \cite{Zolfi:2023bdp}.}, one could argue that the emission of a baby universe effectively makes the black hole younger.
To explore this concept within the framework of JT gravity, one observes that through a specific analytic continuation, the trumpet partition function in JT gravity can be interpreted as a tunneling amplitude between states of different ages \cite{Stanford:2022fdt}.
To be more precise, let us examine the explicit expression of the partition function for the trumpet geometry in JT gravity
\cite{Stanford:2019vob}:
\begin{equation}\label{trup}
	Z_{\text{trumpet}}\left( \beta,a\right) 
    =\frac{\exp\left(-\frac{a^2}{4\beta}\right)}{\sqrt{4\pi \beta}},
\end{equation}
where $\beta$ is the renormalized length of the Euclidean AdS boundary and $a$ is the length of the geodesic boundary (baby universe) of the trumpet.
Let us now consider the analytic continuation $\beta \rightarrow \beta + \text{i}(t - t')$. This transformation allows us to interpret the trumpet 
path integral as a tunneling amplitude from the thermofield double state of age $t$ to the thermofield double state of age $t'$, along with the creation of a baby universe of size $a$, as demonstrated in the following relation:
\begin{figure}[H]
	\centering
	\begin{overpic}
		[width=0.34\textwidth]{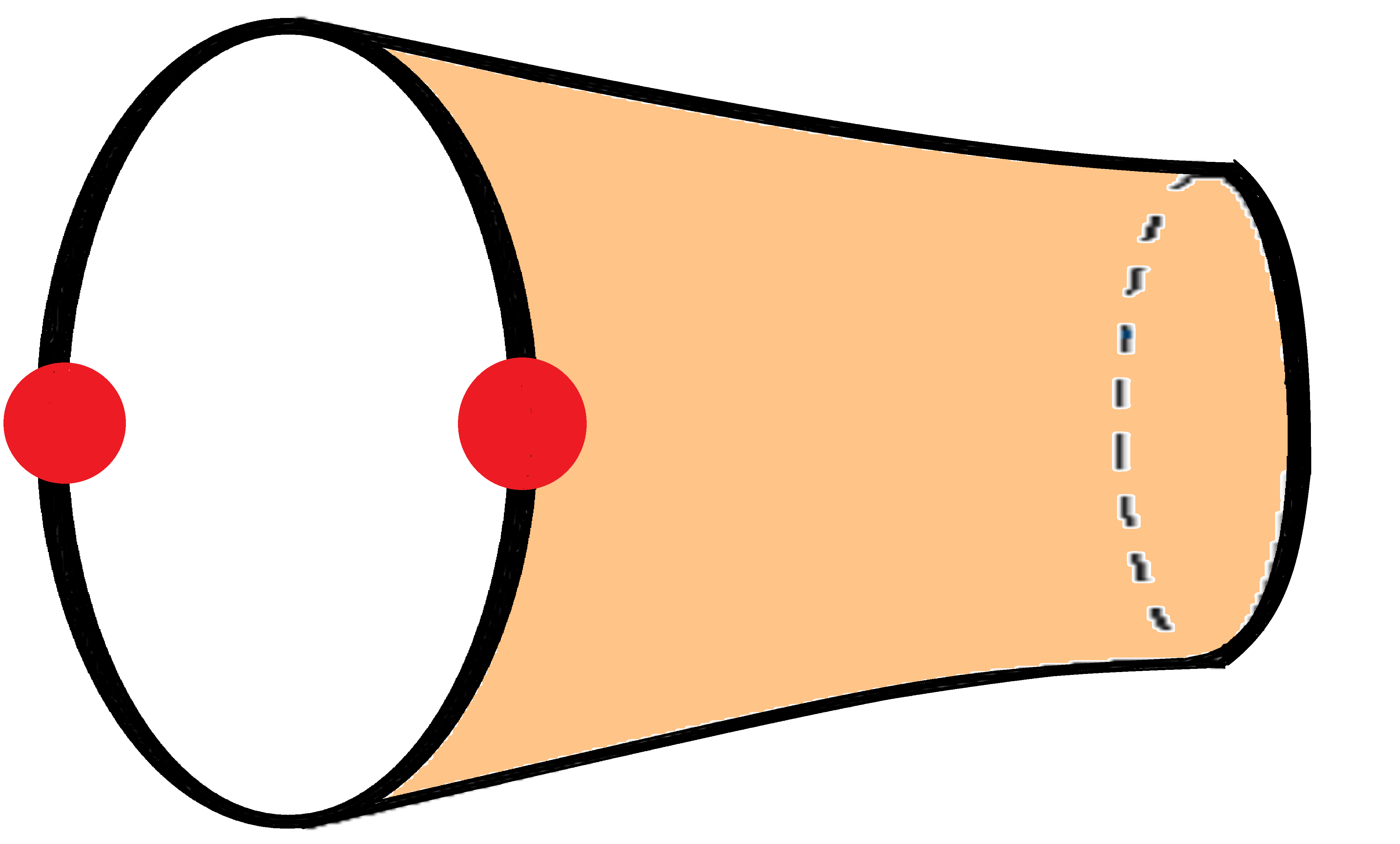}
		\put (98,35) {$\displaystyle a$}	
		\put (11.6,44.6) {$\displaystyle \frac{\beta}{2}+\text{i}t^{\prime}$}
		\put (12,15) {$\displaystyle\frac{\beta}{2}+\text{i}t$}
		\put (-60,28) {$\displaystyle\langle \text{age}{ }~t^{\prime};a|\text{age}~ t\rangle=~$}		
	\end{overpic}
\end{figure}
\vspace{-.9cm}
\begin{align}\label{trpp}
	\qquad\qquad=\frac{\exp\left( -\frac{a^2}{4\left( \beta+\text{i}(t-t^{\prime})\right) }\right) }{\sqrt{4\pi \left(\beta+\text{i}(t-t^{\prime}) \right) }}\approx \frac{\exp\left( \text{i}\frac{a^2}{4 (t-t^{\prime}) }-\beta\frac{a^2}{4 (t-t^{\prime})^{2} }\right) }{\sqrt{4\pi \text{i}(t-t^{\prime})}}.
\end{align}
It is useful to examine this amplitude at a fixed energy, which can be achieved by employing an inverse Laplace transform. To do this, we multiply the amplitude by $e^{\beta E}$ and integrate over 
$\beta$ along an inverse Laplace transform contour. This procedure results in a delta function, yielding:
\begin{equation}
2\sqrt{E} \left( t-t^{\prime}\right) =\pm a.  
\end{equation} 
Discarding the minus sign due to its unphysical implications \cite{Stanford:2022fdt}, we define the wormhole lengths before and after emitting a baby universe as $\ell = 2\sqrt{E}t$ and $\ell' = 2\sqrt{E}t'$, respectively. Then
the above equation can be expressed as:
\begin{equation}
 \ell = \ell' + a, 
\end{equation}
which can be interpreted as the emission of a baby universe making the black hole younger, and equivalently, reducing the length of the wormhole.

It is natural to extend the procedure described above to find the transition amplitude from the path integral of a somewhat complex geometry featuring two or more baby universes and non-zero genus. However, a naive generalization of this procedure leads to a puzzling outcome.

To elaborate, let us consider the partition function for a geometry constructed from a trumpet and a three-holed sphere, which can be utilized to study the emission of two baby universes. The corresponding partition function is:
\begin{figure}[H]
	\centering
	\begin{overpic}
		[width=0.3\textwidth]{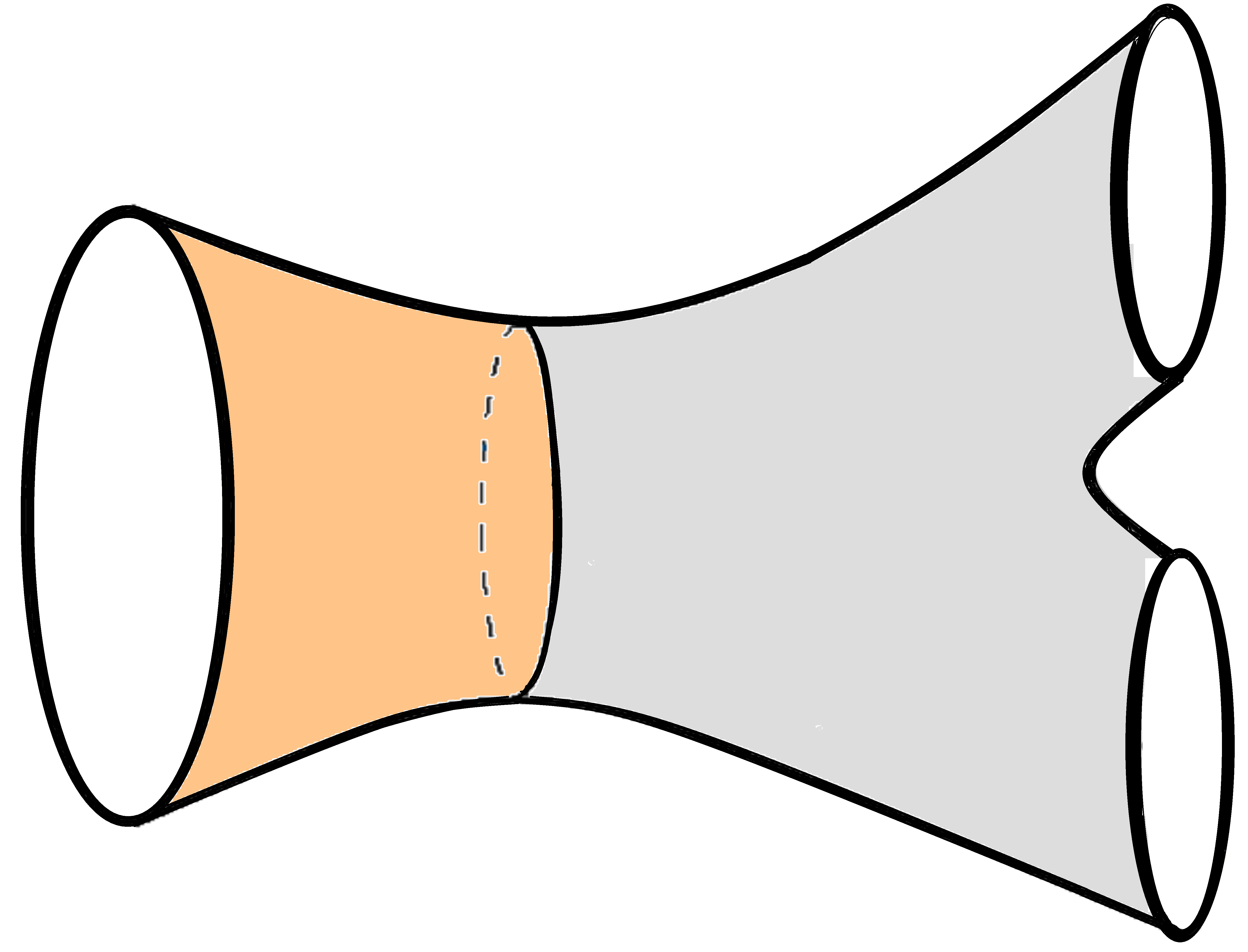}
		\put (-77,30) {$\displaystyle Z_{0}(\beta,a_1,a_2,)	=\qquad\quad~\beta$}	
		\put (48,30) {$\displaystyle a_3$}	
		\put (100,55) {$\displaystyle a_1$}	
		\put (100,10) {$\displaystyle a_2$}	
	\end{overpic}
\end{figure}
\vspace{-1cm}
\begin{align}\label{part3}
\qquad\qquad=e^{S_0\chi }\int_{0}^{\infty}Z_{\text{trumpet}}\left( \beta,a_3\right)V_{0,3}(a_1,a_2,a_3)  a_3\text{d}a_3
    =e^{S_0\chi }\frac{\sqrt{\beta }}{\sqrt{\pi }}.
\end{align}
Here, $V_{0,3}(a_1, a_2, a_3)$ denotes the Weil-Petersson (WP) volume of a geometry with zero genus and three boundaries, where the subscript $0$ indicates that the genus of the geometry is zero. Notably, this volume is known to be independent of the lengths $a_i$: $ V_{0,3}(a_1, a_2, a_3) = 1$\cite{mir}\footnote{The Euler characteristic and the ground state entropy are denoted by $\chi$ and $S_0$, respectively.}.
As a result, the path integral becomes independent of the lengths of the baby universes, as demonstrated in the above expression. This raises challenges in interpreting the tunneling process from a black hole to a white hole in the context of this path integral, rendering it somewhat elusive.

This issue may initially seem to stem from the fact that the WP volume of the three-holed sphere is independent of the geodesic lengths. However, it is worth noting that the problem persists even for geometries with arbitrary numbers of geodesic boundaries and genus, because of the structure of the WP volume.
In fact, for a geometry with $n$ boundaries and 
 genus $g$, this volume is a symmetric polynomial function of the squared geodesic lengths $a_{1}^2, a_{2}^2, \dots, a_{n}^2$, with a degree of $3g - 3 + n$. It can be expressed as follows \cite{mir,mir1}:
\begin{equation}\label{weil}
V_{g,n}(\textbf{a}) = \sum_{|\alpha| \leq 3g - 3 + n} c_{g,n}(\alpha) \prod_{j=1}^{n} \frac{a_j^{2\alpha_j}}{2^{2\alpha_j}(2\alpha_j + 1)!},
\end{equation}
where $\textbf{a}=(a_1,\ldots,a_n)$ represents the geodesic lengths of the boundaries. Although the complexity of the structure may make it tedious to analyze, it is straightforward to see that the aforementioned problem persists in these cases as well.

This paper aims to present a method for 
computing the tunneling amplitude from the 
partition function, building upon the naive 
procedure discussed earlier. This involves focusing 
on a specific region of the moduli space. 
Specifically, we will utilize Kontsevich's approach 
to compute the WP volume in the Airy 
limit using ribbon graphs. Our main observation is that, while all associated graphs must be considered to compute the WP volume of a geometry with a given genus and boundary in the Airy limit, only certain graphs indicate the tunneling process.

The paper is organized as follows: In the next section, we begin by reviewing Kontsevich's approach to computing the volume of the moduli space. We then explore how the tunneling process can occur within specific regions of the moduli space of a three-holed sphere. Section \ref{foor} extends this discussion to other geometries, one involving three baby universes and the other with a single baby universe and a single genus, offering further elaboration on the ideas introduced in Section \ref{tree}. Finally, concluding remarks are presented in Section \ref{con}.

\section{Tunneling in a geometry with constant Weil-Petersson volume}\label{tree}

In this section, we will study the tunneling 
amplitude associated with the emission of two baby 
universes. Building on our discussion in the 
introduction, we need to revisit the partition 
function of geometry with one asymptotic boundary and two geodesic 
boundaries. To achieve this, we will employ 
Kontsevich's approach to compute the WP
volume in the Airy limit, which involves 
calculating certain ribbon graphs. Utilizing these 
results, we can determine the contribution of each 
graph to the partition function. While it is essential to consider the contributions of all graphs to obtain the expected partition function, only some may yield a physically acceptable tunneling amplitude. To proceed, the next subsection briefly reviews Kontsevich's approach to computing the corresponding volume.

\subsection{Kontsevich's decomposition of moduli space}\label{tori}	

To review Kontsevich's approach to computing the WP volume, we consider surfaces of genus 
$g$ with 
$n$ geodesic boundaries. In a specific case known as the Airy limit, or thin-strip limit, these boundaries extend indefinitely while the overall area of the surfaces remains fixed.  
Under this condition, the hyperbolic surfaces transition into what are known as ribbon graphs \cite{kon, do}. 

In practice, instead of 
depicting the entire surface, we represent it using 
a trivalent graph. A trivalent graph is 
characterized by each vertex connecting exactly 
three edges. In this representation, lengths are assigned to the edges, while the bulk geometry of the corresponding surface is not explicitly shown.
The number of edges 
$E$ and vertices 
$V $ in such a graph is given by the following relations:
\begin{equation}
 E = 6g - 6 + 3n,\qquad\qquad V = 4g - 4 + 2n,    
\end{equation}
yielding 
\begin{equation}
  V - E = 2 - 2g - n.  
\end{equation}
The above relation is essentially the Euler 
characteristic formula, which highlights a 
fundamental property of the graph associated with 
the ribbon graph structure.
\begin{figure}[h]
	\centering
	\begin{overpic}
		[width=0.7\textwidth]{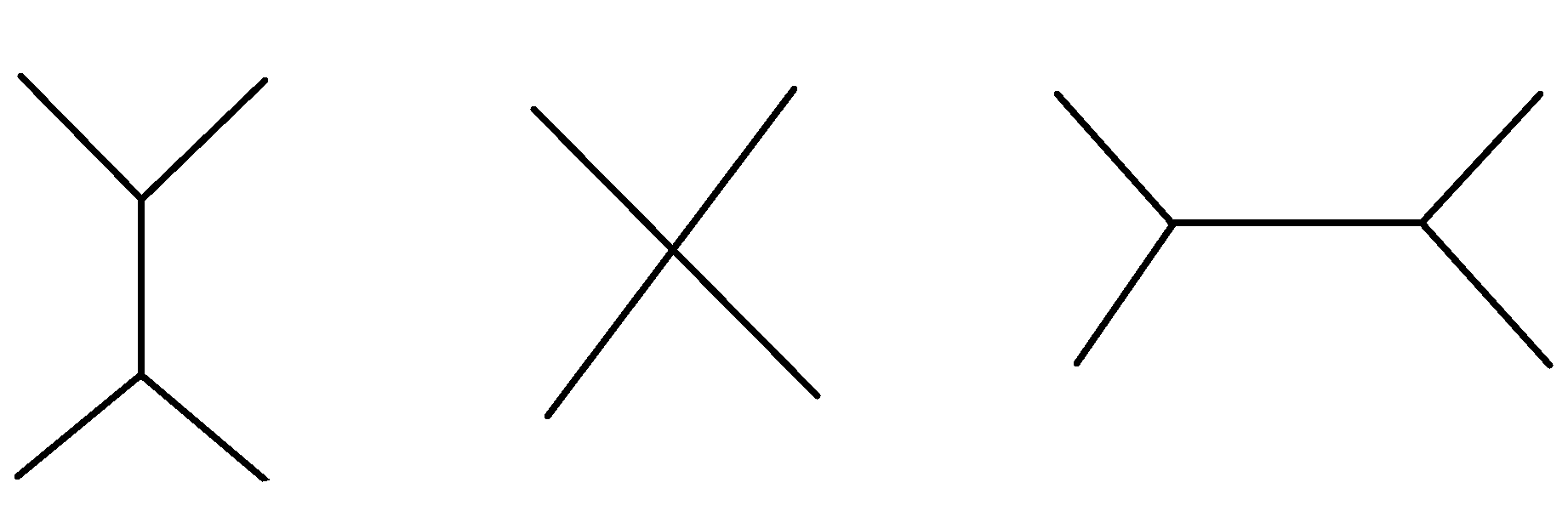}
		\put (7,23) {$\displaystyle w_2$}	
		\put (10,13) {$\displaystyle e$}	
			\put (7,5) {$\displaystyle w_1$}	
		\put (45,17) {$\displaystyle w$}	
		\put (69,18) {$\displaystyle w'_1$}	
		\put (93,18) {$\displaystyle w'_2$}	
	\put (83,20) {$\displaystyle e'$}	
	\end{overpic}
	\caption{Illustration of the Whitehead collapse process on a  graph  $G$, where collapsing an edge  $e$  merges vertices  $w_1$  and  $w_2$  into a new vertex $w'$. In the resulting graph $ G' $, a new edge $ e'$  is added in a different direction, forming a trivalent graph.
	}	
	\label{trsg}
\end{figure}

Considering the thin-strip limit offers a useful simplification for analyzing the moduli space, where the WP volume \eqref{weil} is replaced by the Airy volume, defined as:
\begin{equation}
 V^{\text{Airy}}_{g,n}(\textbf{a})=\lim_{\Lambda\rightarrow\infty}\Lambda^{3-3g-n} V_{g,n}(\Lambda\textbf{a}).  
\end{equation}
Under this approximation, the moduli space can be expressed as a sum over trivalent ribbon graphs. In this context, we also incorporate an integral that accounts for the lengths of the edges that form these graphs, while ensuring that the lengths of the boundaries are fixed at specified values $a_i$. Consequently, one arrives at the following expression for the volume in the Airy limit \cite{kon,do}:
\begin{equation}\label{moduli}
	V^{\text{Airy}}_{g,n}(\textbf{a})=\sum_{\Gamma\in\Gamma_{g,n}}\frac{2^{2g-2+n}}{\left|\text{Aut}\left(\Gamma_{g,n} \right)  \right| }\prod_{k=1}^{E}\int_{0}^{\infty}\text{d}y_{k}\prod_{i=1}^{n}\delta\left( a_i-\sum_{k=1}^{n}n^{i}_{k}y_{k}\right). 
\end{equation}
Here, $\Gamma_{g,n}$ denotes the set of trivalent ribbon graphs of genus $g$ and $n$ boundaries, known as Kontsevich graphs.  
In the above expression $y_k$ is the length of edge $k$, and
$n_i\in \left\lbrace 0,1,2 \right\rbrace $
denotes the number of sides of edge $k$ that belong to boundary $i$.

It is also useful to perform the Laplace transform of the expression \eqref{moduli} to obtain:
\begin{align}\label{laplace}
	\tilde{	V}^{\text{Airy}}_{g,n}(\textbf{z})&\equiv\int_{0}^{\infty}\prod_{i=1}^{n}\text{d}		a_{i}e^{-z_i a_i}V^{\text{Airy}}_{g,n}(\textbf{a})\nonumber\\
	&=\sum_{\Gamma_{g,n}}\frac{2^{2g-2+n}}{\left|\text{Aut}\left(\Gamma_{g,n} \right)  \right| }\prod_{k=1}^{E}\frac{1}{z_{l\left( k\right) }+z_{r\left( k\right) }}, 
\end{align}
that looks simpler and more intuitive for the computation of the WP volume in the Airy limit. Here the $l\left( k\right)\in\left\lbrace 1\dots n \right\rbrace $  index labels which boundary of the Riemann surface
the left side of the ribbon belongs to. Similarly, $r(k)$ labels which boundary the right side of the
ribbon belongs to.

To enumerate the graphs, it is useful to note that all orientable graphs for fixed 
$(g,n)$  can be derived from a single graph through repeated application of the cross operation (or Whitehead collapse).
Starting with a trivalent graph $G$ that contains an edge $e$ and two vertices $w_1$ and 
$w_2$, one can obtain another trivalent  graph $G'$ by merging the vertices $w_1$ and 
$w_2$ into a single vertex $w$. Subsequently, one can expand in a different direction to create a new edge $e'$ and introduce new vertices $w'_{1}$ and 
$w'_{2}$, as illustrated in figure \ref{trsg}. This entire process is referred to as a Whitehead collapse, which results in a new graph while preserving certain important structural properties \cite{wc, penner}.

\subsection{Geometry with two baby universes}
\begin{figure}[h]
	\centering
	\begin{overpic}
		[width=0.95\textwidth]{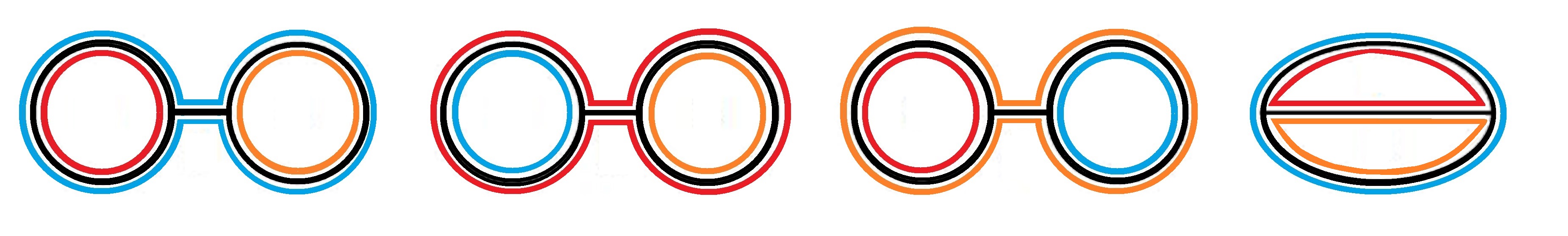}
		\put (12,0) {$\displaystyle \text{I}$}	
		\put (38,0) {$\displaystyle \text{II}$}	
		\put (63,0) {$\displaystyle \text{III}$}	
		\put (86,0) {$\displaystyle \text{IV}$}	
	\end{overpic}
	\caption{
The diagram shows the four trivalent  Kontsevich graphs of type (0, 3). The colored curves denote the boundaries of a three-holed sphere: the red curve corresponds to the boundary with length $a_1$, the orange curve to the boundary with length $a_2$, and the blue curve to the boundary with length $a_3$.}	
	\label{trg}
\end{figure}

In this subsection, the WP volume of a three-holed sphere with geodesic boundaries denoted by $ a_1, a_2, a_3 $ in the Airy limit will be computed using Kontsevich's approach. To begin, all possible ribbon graphs associated with the three-holed sphere must be identified. It is relatively straightforward to show that, in this case, the corresponding graphs consist of three edges and two vertices, leading to the consideration of four specific graphs, as depicted in figure \ref{trg}.
Using the formula \eqref{moduli}, the contribution of the first Kontsevich graph to the volume
$V_{0,3}(a_1, a_2, a_3)$ is:
\begin{align}\label{air}
	V_{0,3}^{\text{I}}(a_1,a_2,a_3)&=2\int_{0}^{\infty}\text{d}y_{1}\text{d}y_{2}\text{d}y_{3}\delta\left( y_1+y_2+2y_3-a_3\right)\delta\left(y_1- a_1\right) \delta\left( y_2-a_2\right)
	\nonumber\\
	&=2\int_{0}^{\infty}\text{d}y_{3}\delta\left( a_1+a_2+2y_3-a_3\right)= \theta \left(a_3-a_1-a_2\right).
\end{align}
Similarly, the contributions from the second and third Kontsevich graphs can be computed, as they are essentially obtained through permutations of the first graph:
\begin{align}\label{gif}
	V_{0,3}^{\text{II}}(a_1,a_2,a_3)= \theta \left(a_1-a_3-a_2\right),
\qquad\qquad
	V_{0,3}^{\text{III}}(a_1,a_2,a_3)= \theta \left(a_2-a_1-a_3\right).
\end{align}
The last Kontsevich graph in figure \ref{trg}, which is symmetric under the permutation of the geodesics, yields the following contribution:
\begin{align}
	V_{0,3}^{\text{IV}}(a_1,a_2,a_3)&=2\int_{0}^{\infty}\text{d}y_{1}\text{d}y_{2}\text{d}y_{3}\delta\left( y_1+y_2-a_3\right)\delta\left(y_1+y_3- a_1\right) \delta\left( y_2+y_3-a_2\right)
	\nonumber\\
	&= \theta \left(a_1+a_2-a_3\right)\theta \left(a_1-a_2+a_3\right)\theta \left(a_3-a_1+a_2\right),
\end{align}
which is also symmetric under the permutation of the geodesics. It follows from the sum of all contributions that \begin{align} V_{0,3}(a_3, a_1, a_2) = \sum_{i=\text{I}}^{\text{IV}} V_{0,3}^{i}(a_3, a_1, a_2) = 1, \end{align} as expected.

It is also useful to compute the contribution of each Kontsevich graph to the path integral
given in \eqref{part3}. To do so, one may 
attach each Kontsevich graph to the trumpet along the geodesic $a_3$.  For the first graph, the contribution is given by:
\begin{align}\label{3fw1}	\mathcal{Z}_{0}^{\text{I}}	(\beta,a_1,a_2)&=e^{S_0\chi }\int_{0}^{\infty}Z_{\text{trumpet}}\left( \beta,a_3\right)V_{0,3}^{\text{I}}(a_1,a_2,a_3)  a_3\text{d}a_3
	=e^{S_0\chi }\frac{\sqrt{\beta } }{\sqrt{\pi }}\exp\left( -\frac{\left(a_1+a_2\right){}^2}{4 \beta }\right)\,.
\end{align}
Similarly, the contributions from the other Kontsevich graphs are:
\begin{align}
&\mathcal{Z}_{0}^{\text{II}}(\beta,a_1,a_2)
=e^{S_0\chi }\frac{\sqrt{\beta }  }{ \sqrt{\pi }} \left(1-\exp\left( -\frac{\left(a_1-a_2\right){}^2}{4 \beta }\right) \right)\theta(a_1-a_2),	
\nonumber \\
&\mathcal{Z}_{0}^{\text{III}}(\beta,a_1,a_2)
=e^{S_0\chi }\frac{\sqrt{\beta }  }{ \sqrt{\pi }} \left(1-\exp\left( -\frac{\left(a_1-a_2\right){}^2}{4 \beta }\right) \right)\theta(a_2-a_1),	\nonumber \\
&\mathcal{Z}_{0}^{\text{IV}}(\beta,a_1,a_2)
=e^{S_0\chi }\frac{\sqrt{\beta } }{ \sqrt{\pi }} \left( \exp\left( -\frac{\left(a_1-a_2\right){}^2}{4 \beta }\right)  -\exp\left( -\frac{\left(a_1+a_2\right){}^2}{4 \beta }\right) \right)\,,
\end{align}
It is evident that their sum yields the partition function in \eqref{part3}.

As previously mentioned, the partition function of the trumpet can be interpreted as the tunneling amplitude from the thermofield double state of age $t$ to the thermofield double state of age $t'$, along with a baby universe of size $a$, after replacing $\beta$ with $\beta + \text{i}(t - t')$.Naturally, this interpretation can be extended to the case where two baby universes are emitted. In this scenario, the tunneling amplitude associated with Kontsevich graph I is given by:
\begin{align}\label{tqrp}
	\langle \text{age}{ }~t^{\prime};a_1,a_2|\text{age}~ t\rangle_{\text{I}}&=e^{S_0\chi }\frac{\sqrt{\beta+\text{i}(t-t^{\prime}) } }{\sqrt{\pi }}\exp\left( -\frac{\left(a_1+a_2\right){}^2}{ 4\left( \beta+\text{i}(t-t^{\prime})\right) }\right) \nonumber\\& \approx e^{S_0\chi }\frac{\sqrt{\text{i}(t-t^{\prime}) } }{\sqrt{\pi }}\exp\left( \text{i}\frac{\left(a_1+a_2\right){}^2}{4 (t-t^{\prime}) }-\beta\frac{\left(a_1+a_2\right){}^2}{4 (t-t^{\prime})^{2} }\right),
\end{align}
where the second line corresponds to small $\beta$ limit with respect to $t-t^\prime$. Moreover, to obtain the  amplitude at fixed energy, one may perform an inverse Laplace transformation over $\beta$
to arrives at: 
\begin{align}\label{tqbry}
e^{S_0\chi }\frac{\sqrt{\text{i}(t-t^{\prime}) } }{\sqrt{\pi }}\exp\left( \text{i}\frac{\left(a_1+a_2\right)^2}{4 (t-t^{\prime}) }\right)\delta\left(E-\frac{\left(a_1+a_2\right)^2}{4 (t-t^{\prime})^{2} } \right).
\end{align}
From the delta function, one obtains:
\begin{align}\label{tqbury}
2\sqrt{E}(t - t^{\prime}) = a_1 + a_2,
\end{align}
indicating that after the emission of baby universes of sizes $a_1$ and $a_2$, the black hole becomes younger, similar to the behavior observed for the trumpet.
Applying the same calculation to the Kontsevich graph 
II (or III) leads to
\begin{align}\label{tqry}
-e^{S_0\chi }\frac{\sqrt{\text{i}(t-t^{\prime}) } }{\sqrt{\pi }}\exp\left( \text{i}\frac{\left(a_1-a_2\right)^2}{4 (t-t^{\prime}) }\right)	\delta\left(E-\frac{\left(a_1-a_2\right)^2}{4 (t-t^{\prime})^{2} } \right),
\end{align}
from which one obtains the following constraint:
\begin{align}\label{aqbry}
	2\sqrt{E}(t-t^{\prime})=\pm\left(a_1-a_2\right).
\end{align}
From this expression, it is evident that it cannot be interpreted as a black hole emitting two baby universes. To illustrate this, one can set $a_1 = a_2$ without loss of generality. Under this assumption, the expression \eqref{tqry} becomes zero. Intuitively, one would expect that this assumption—that the two baby universes have the same size—should not lead to a vanishing tunneling amplitude. Therefore, it can be concluded that interpreting the quantity 
$\mathcal{Z}_{0}^{\text{II,III}} (\beta, a_1, a_2)$ as a tunneling amplitude is incorrect. 
From the expression of $\mathcal{Z}_{0}^{\text{IV}}(\beta, a_1, a_2)$, it is clear that this graph does not introduce any new constraints.


To conclude, the tunneling process of emitting baby universes, which leads to younger black holes, is evident in certain parts of moduli space, particularly in graphs I and IV. However, the sum of their contributions, i.e., the expressions
$\mathcal{Z}_{0}^{\text{I}}(\beta, a_1, a_2)$ and $\mathcal{Z}_{0}^{\text{IV}}(\beta, a_1, a_2)$, does not show this effect. In contrast, the combined contributions of graphs II, III, and IV do. 

The restriction of moduli space has been assumed in the considerations of \cite{Stanford:2022fdt}. The authors introduce a delta function into the path integral of a three-holed sphere, focusing on a specific region of the moduli space associated with firewall geometries. This delta function selects geometries containing wormholes of a particular length to assess the probability of encountering a firewall.

\section{Tunneling in  geometries with nontrivial Weil-Petersson volumes }\label{foor}
To further investigate this approach to computing the tunneling amplitude using the path integral and WP volume, this section examines a black hole emitting three baby universes, corresponding to a four-holed sphere geometry. Additionally, a single-genus geometry with one baby universe is considered, where the genus of the surface introduces a more intricate dependence on the WP volume. In these cases, unlike the three-holed sphere, for which the WP volume remains constant, the WP volumes of these surfaces depend on the parameters that define the geodesic lengths.
\subsection{Geometry with three baby universes}
The WP volume of a four-holed sphere is given by \cite{mir}:
\begin{align}\label{volf}
V_{0,4}(\textbf{a}) = 2\pi^2 + \frac{1}{2} \sum_{i=1}^{4} a_i^2,
\end{align}
 However, the same puzzle persists: the tunneling amplitude does not explicitly reveal a relationship between the effective age of the black hole and the sizes of the emitted baby universes. To make this more concrete, consider the JT gravity path integral of a geometry with one asymptotic boundary and three geodesic boundaries: 
\begin{figure}[H]
	\centering
	\begin{overpic}
		[width=0.25\textwidth]{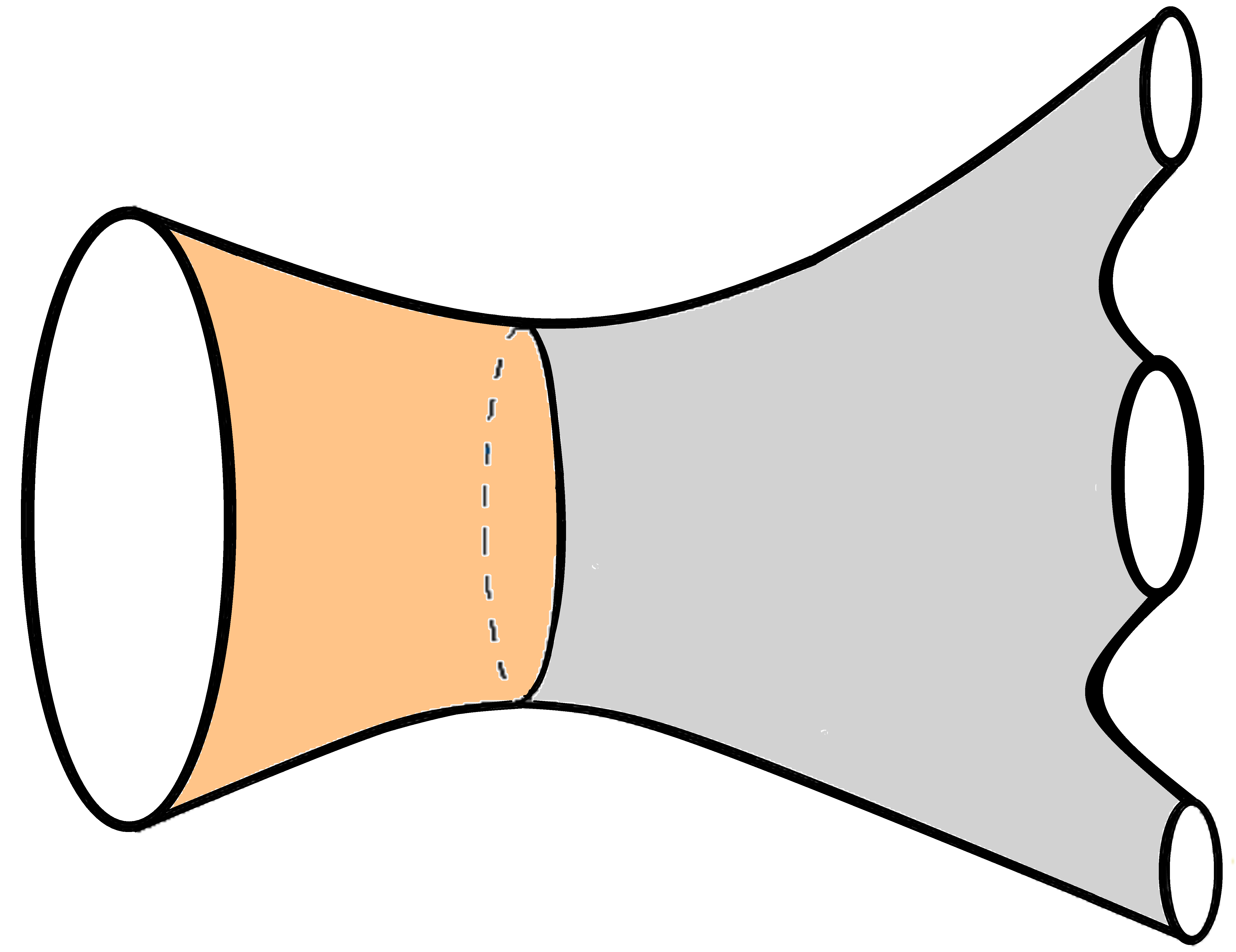}
		\put (10,30) {$\displaystyle \beta$}	
		\put (48,30) {$\displaystyle a_4$}	
		\put (100,73) {$\displaystyle a_3$}	
		\put (100,40) {$\displaystyle a_2$}
			\put (101,5) {$\displaystyle a_1$}
 	\put (-90,35) {$\displaystyle Z_{0}(\beta,a_1,a_2,a_3)=$}     
	\end{overpic}
\end{figure}
\vspace{-1cm}
\hspace{-2cm}
\begin{align}\label{part4}
\qquad\qquad	&=e^{S_0\chi }\int_{0}^{\infty}Z_{\text{trumpet}}\left( \beta,a_4\right)V_{0,4}(a_1,a_2,a_3,a_4)  a_4\text{d}a_4
\nonumber\\ &=e^{S_0\chi }\frac{\sqrt{\beta } }{2 \sqrt{\pi }}\left(a_1^2+a_2^2+a_3^2+4 \left(\beta +\pi ^2\right)\right).
\end{align}
As can be seen from this expression, the transition amplitude from the black hole to the white hole is not immediately apparent in the above path integral. The goal is to extract a portion of the moduli space that reveals this property.

To proceed, following the analysis in the previous section, we first need to compute the WP volume of the four-holed sphere in the Airy limit using Kontsevich's approach. This involves calculating the contributions from specific six-ribbon graphs, assuming 
$a_4>a_1+a_2+a_3$  for simplicity, as detailed in Appendix \ref{iii}.
Utilizing the results from this appendix, we will examine the path integral \eqref{part4} associated with each ribbon graph. From there, we will derive the transition amplitude and investigate the relationship between the age of the black hole and the lengths of the emitted baby universes.

To begin, we compute the portion of the path integral \eqref{part4} associated with the Kontsevich graph, which incorporates the WP volume \eqref{lapry}. This contribution is given by:
\begin{align}
\mathcal{Z}_{0}^{\text{I}}	(\beta,a_1,a_2,a_3)	&=e^{S_0\chi }\int_{0}^{\infty}Z_{\text{trumpet}}\left( \beta,a_4\right)V_{0,4}^{\text{I}}(a_1,a_2,a_3,a_4)  a_4\text{d}a_4\nonumber\\&=e^{S_0\chi }c_1\bigg\lbrace\frac{2 \beta ^{3/2}  }{\sqrt{\pi }}\exp\left( -\frac{\left(a_1+a_2+a_3\right){}^2}{4 \beta }\right) -\beta\left(a_1+a_2+a_3\right) \text{erfc}\left(\frac{a_1+a_2+a_3}{2 \sqrt{\beta }}\right)\bigg\rbrace.
\end{align}

The tunneling amplitude from the thermofield double state of age $t$ to the thermofield double state with age $t'$, plus baby universes of sizes $a_1, a_2$, and $a_3$, can be obtained via the analytic continuation
$\beta \rightarrow \beta + \text{i}(t - t') $ as following:
\begin{align}
	\langle \text{age}{ }~t^{\prime};\textbf{a}|\text{age}~ t\rangle_{\text{I}}&
=e^{S_0\chi }c_1\bigg\lbrace\frac{2 \left(\beta+\text{i}(t-t^{\prime}) \right) ^{3/2}  }{\sqrt{\pi }}\exp\left( -\frac{\left(a_1+a_2+a_3\right)^2}{4\left( \beta+\text{i}(t-t^{\prime})\right)  }\right) \nonumber\\&\quad-\left(a_1+a_2+a_3\right) \left(\beta+\text{i}(t-t^{\prime}) \right) 
	\text{erfc}\left(\frac{a_1+a_2+a_3}{2 \sqrt{\beta+\text{i}(t-t^{\prime}) }}\right)\bigg\rbrace\,.
\end{align}
In the low $\beta$ limit, this expression simplifies to:
 \begin{align}\label{trp}  
 \langle \text{age}{ }~t^{\prime};\textbf{a}|\text{age}~ t\rangle_{\text{I}}
&\approx e^{S_0\chi }c_1\bigg\lbrace\frac{2 \left(\text{i}(t-t^{\prime}) \right) ^{3/2}  }{\sqrt{\pi }}\exp\left( \text{i}\frac{\left(a_1+a_2+a_3\right)^2}{4 (t-t^{\prime})}  -\beta\frac{\left(a_1+a_2+a_3\right)^2}{4 (t-t^{\prime})^{2} }\right) \nonumber\\&\quad-\left(a_1+a_2+a_3\right) \text{i}(t-t^{\prime})  
		\text{erfc}\left(\frac{a_1+a_2+a_3}{2 \sqrt{\text{i}(t-t^{\prime}) }}-\beta\frac{a_1+a_2+a_3}{4 \left( \text{i}(t-t^{\prime}) \right)^{\frac{3}{2}} }\right)\bigg\rbrace.		
\end{align}
To obtain the amplitude at fixed energy, one may perform an inverse Laplace transformation over 
$\beta$. For the term involving the complementary error function (erfc), the result is:
\begin{align}\label{tqdo}
\frac{\text{i}}{\pi  E} \exp \left(2 \text{i} E \left(t-t'\right)\frac{
   \left(\left(a_1+a_2+a_3\right){}^2-2 E
   \left(t-t'\right)^2\right)}{\left(a_1+a_2+a_3\right){}^2}\right)\,,
\end{align}
whereas, the exponential term produces the familiar delta function:
\begin{align}\label{bqbry}
\frac{\sqrt{\text{i}(t-t^{\prime}) } }{\sqrt{\pi }}\exp\left( \text{i}\frac{\left(a_1+a_2+a_3\right)^2}{4 (t-t^{\prime}) }\right)\delta\left(E-\frac{\left(a_1+a_2+a_3\right)^2}{4 (t-t^{\prime})^{2} } \right).
\end{align}

It is also straightforward to compute the contributions from the ribbon graphs outlined in the Appendix, which can establish a relationship between the effective age of the black hole and the length of the baby universe. The contributions from the WP volume of the Kontsevich graphs \eqref{ldry}, \eqref{iiia}, and \eqref{iiib}, under the assumption that $a_4>a_1+a_2+a_3$ for the latter two graphs, are given respectively by:   
\begin{align}
	\mathcal{Z}_{0}^{\text{II}}	(\beta,a_1,a_2,a_3)	
    &=e^{S_0\chi }c_2\beta \text{erfc}\left(\frac{a_1+a_2+a_3}{2 \sqrt{\beta }}\right)\left(a_1+a_2+a_3\right), \nonumber
    \\
	\mathcal{Z}^{\text{IIIA}}_{0}	(\beta,a_1,a_2,a_3)	
    &=e^{S_0\chi }c_3\frac{\beta }{4}   \bigg\lbrace\frac{2 \sqrt{\beta } }{\sqrt{\pi }}\left(\exp\left( -\frac{\left(-a_1+a_2+a_3\right){}^2}{4 \beta
	}\right) -\exp\left(-\frac{\left(a_1+a_2+a_3\right){}^2}{4 \beta }\right) \right)\nonumber\\&\quad\quad\quad+\left(a_1-a_2+a_3\right)
	\left(\text{erf}\left(\frac{-a_1+a_2+a_3}{2 \sqrt{\beta }}\right)-\text{erf}\left(\frac{a_1+a_2+a_3}{2 \sqrt{\beta
	}}\right)\right)\bigg\rbrace, \nonumber\\
	\mathcal{Z}_{0}^{\text{IIIB}}	(\beta,a_1,a_2,a_3)	
    &=e^{S_0\chi }c_3\beta    \bigg\lbrace\frac{2 \sqrt{\beta } }{\sqrt{\pi }}\left(\exp\left( -\frac{a_1^2}{4 \beta }\right) -\exp\left( -\frac{\left(a_1-a_2+a_3\right){}^2}{4 \beta
	}\right) \right)\nonumber\\&\qquad\qquad\qquad+a_1 \left(\text{erf}\left(\frac{a_1}{2 \sqrt{\beta }}\right)-\text{erf}\left(\frac{a_1-a_2+a_3}{2
		\sqrt{\beta }}\right)\right)\bigg\rbrace.
\end{align}
As observed with the first graph, the exponential terms yield delta functions, while the terms involving the complementary error function (erfc) do not impose any constraints on the parameters. The corresponding delta functions are:
\begin{align}\label{del2}
&\frac{\sqrt{\text{i}(t-t^{\prime}) } }{\sqrt{\pi }}\exp\left( \text{i}\frac{\left(a_2-a_1+a_3\right)^2}{4 (t-t^{\prime}) }\right)\delta\left(E-\frac{\left(a_2-a_1+a_3\right)^2}{4 (t-t^{\prime})^{2} } \right),\nonumber\\
&\frac{\sqrt{\text{i}(t-t^{\prime}) } }{\sqrt{\pi }}\exp\left( \text{i}\frac{\left(a_1-a_2+a_3\right)^2}{4 (t-t^{\prime}) }\right)\delta\left(E-\frac{\left(a_1-a_2+a_3\right)^2}{4 (t-t^{\prime})^{2} } \right)\,.
\end{align}
From the relations \eqref{bqbry} and \eqref{del2}, the following constraints are obtained:
\begin{align}
&2\sqrt{E}(t-t^{\prime})=\left(a_1+a_2+a_3\right),\nonumber\\
&2\sqrt{E}(t-t^{\prime})=\pm\left(a_1-a_2+a_3\right),\nonumber\\
&2\sqrt{E}(t-t^{\prime})=\pm\left(a_2-a_1+a_3\right)\,.\label{g1}
\end{align}
From the above expressions, it can be observed that graphs I and IIIA exhibit the notion of a tunneling process. 

If we relax the assumption $a_4>a_1+a_2+a_3$ for the  Kontsevich graphs IV and V, they can also be interpreted as representing a tunneling process. This interpretation arises due to the presence of the term:
\begin{align}
\beta^2 \exp\left(-\frac{\left(a_1 + a_2 + a_3\right)^2}{4 \beta }\right),
\end{align}
in their contribution to the path integral \eqref{part4} in the Airy limit.

\subsection{ Single-genus geometry with a baby universe}\label{gen}

It is also insightful to consider the tunneling amplitude for the simplest geometry with a single genus. This includes a Riemann surface of genus one with two boundaries, which, when attached to a trumpet, provides the one-loop correction to the partition function for the emission of a single baby universe, as discussed in the introduction. The motivation for performing this computation is to demonstrate that the issue encountered in the previous section—specifically, that only certain graphs contribute to the amplitude—also arises in this case.

\begin{figure}[H]
	\centering
	\begin{overpic}
		[width=.95\textwidth]{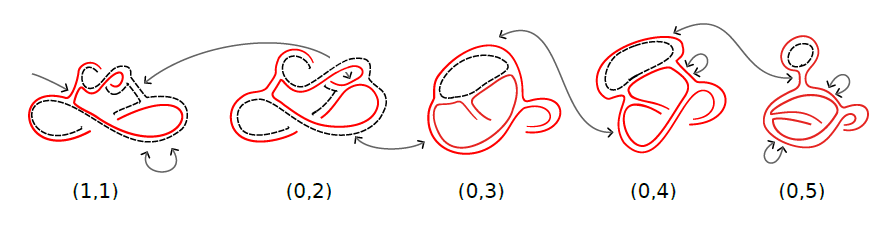}  
	\end{overpic}
	\caption{The nine Kontsevich graphs with two boundaries and genus one consist of these five, along with four additional graphs obtained by interchanging the solid (red) and dashed (black) lines in the last four graphs. The dashed (black) lines correspond to boundary 1, while the solid (red) lines correspond to boundary 2. The gray lines with arrows indicate the effect of applying a cross-operation to a given edge.   }	
	\label{stan}
\end{figure}

The WP volume of the corresponding 
geometry is\cite{mir1}:
\begin{equation}\label{volairy}
	V_{1,2}\left( a_{1},a_2\right) = \frac{1}{192}\left(4\pi^{2}+a_{1}^2+a_{2}^2 \right)\left(12\pi^{2}+a_{1}^2+a_{2}^2 \right), 
\end{equation}
which results in  the partition function: 
\begin{figure}[H]
	\centering
	\begin{overpic}
		[width=0.33\textwidth]{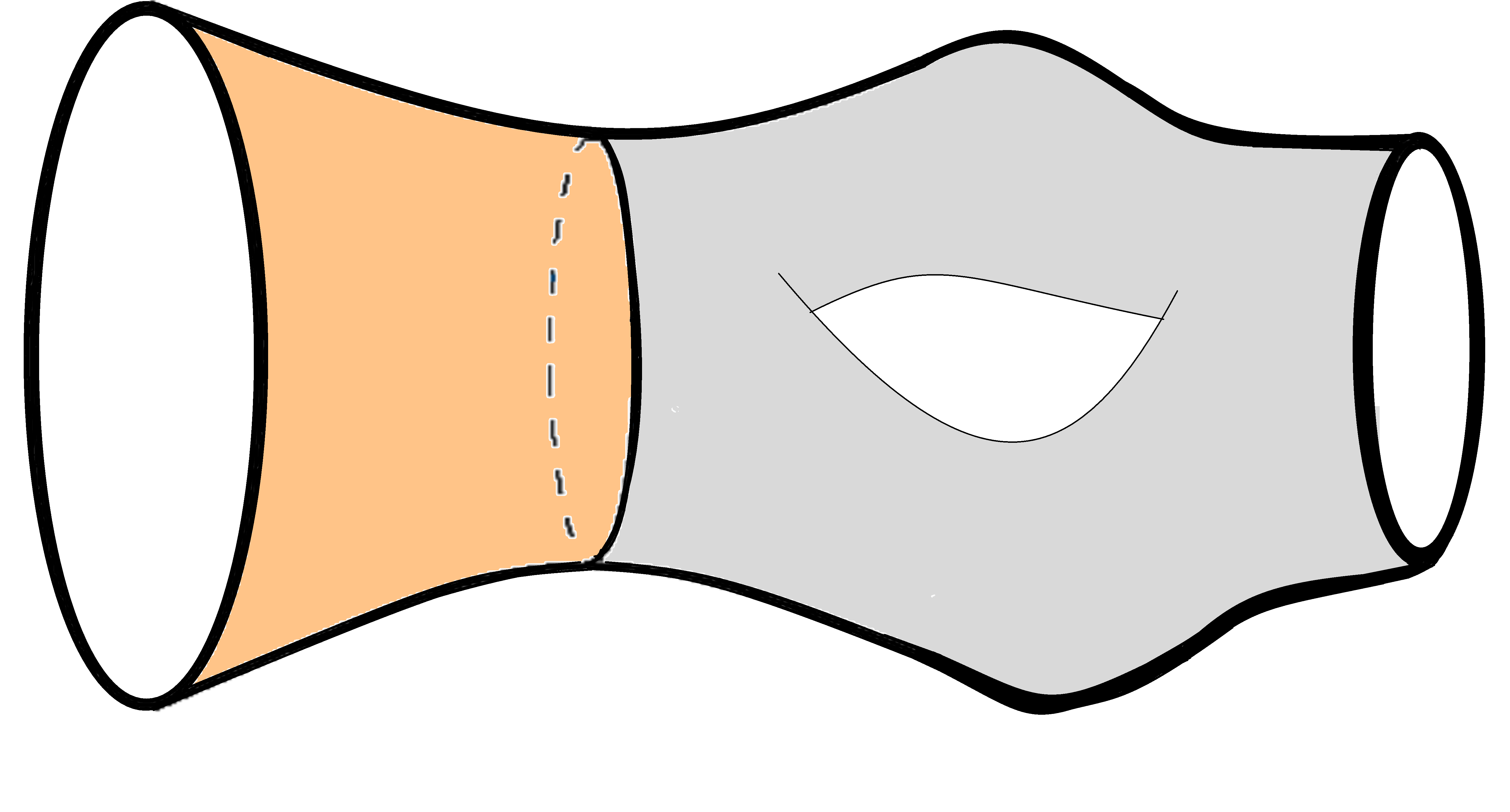}
		\put (10,30) {$\displaystyle \beta$}	
		\put (45,30) {$\displaystyle a_2$}		
		\put (100,30) {$\displaystyle a_1$}
 	\put (-40,30) {$\displaystyle Z_{1}(\beta,a_1)=$}     
	\end{overpic}
\end{figure}
\vspace{-1cm}
\hspace{-2cm}
\begin{align}\label{rg}
\qquad\qquad	&=e^{S_0\chi }\int_{0}^{\infty}Z_{\text{trumpet}}\left( \beta,a_4\right)V_{1,2}(a_1,a_2)  a_2\text{d}a_2
\nonumber\\ &=e^{S_0\chi }\frac{\sqrt{\beta } }{192 \sqrt{\pi }}\left(a_{1}^4+8 a_{1}^2 \left(\beta +2 \pi ^2\right)+16 \left(2 \beta ^2+4 \pi
   ^2 \beta +3 \pi ^4\right)\right).
\end{align}
To compute the WP volume $V_{1,2}$ in the Airy limit using Kontsevich's approach,  one must consider nine distinct ribbon graphs, as shown in figure \eqref{stan}. This figure is included with permission from \cite{Saad:2022kfe}. The graphs are labeled by the number of propagators, and their contributions to $\tilde{	V}^{\text{Airy}}_{1,2}(z_1,z_2)$ are:
 \begin{align}\label{hr}
&\tilde{V}_{1,2}^{\text{(0,5)}}(a_1,a_2)=\frac{1}{8z_2^{5}(z_1+z_2)},
\qquad V_{1,2}^{\text{(0,4)}}(a_1,a_2)=\frac{1}{8z_2^{4}(z_1+z_2)^2}
\nonumber\\
&\tilde{V}_{1,2}^{\text{(0,3)}}(a_1,a_2)=\frac{1}{6z_2^{3}(z_1+z_2)^3}
,
\qquad\tilde{V}_{1,2}^{\text{(0,2)}}(a_1,a_2)=\frac{1}{4z_2^{2}(z_1+z_2)^4},\nonumber\\
& V_{1,2}^{\text{(1,1)}}(a_1,a_2)=\frac{1}{2z_1z_2(z_1+z_2)^4}
\end{align}
 The WP volume associated with each of these graphs is: 
\begin{align}\label{hr}
&V_{1,2}^{\text{(0,5)}}(a_1,a_2)
=\frac{1}{192}\left(a_2-a_1\right)^4\theta\left(a_2-a_1\right),
\nonumber\\&V_{1,2}^{\text{(0,4)}}(a_1,a_2)
=\frac{1}{48} \left(a_2-a_1\right)^3 a_1  \theta \left(a_2-a_1\right),
\nonumber\\
&V_{1,2}^{\text{(0,3)}}(a_1,a_2)
=\frac{1}{24} \left(a_2-a_1\right)^2 a_1^{2} \theta \left(a_2-a_1\right),
\nonumber\\& V_{1,2}^{\text{(0,2)}}(a_1,a_2)
=\frac{1}{24} \left(a_2-a_1\right) a_1^{3} \theta \left(a_2-a_1\right),\nonumber\\
& V_{1,2}^{\text{(1,1)}}(a_1,a_2)
=\frac{1}{48} \left(a_1^{4} \theta \left(a_2-a_1\right)+a_2^{4} \theta \left(a_1-a_2\right)\right).
\end{align}
The contributions of other graphs to the volume $V_{1,2}^{\text{Airy}}(a_1,a_2)$ can be determined by swapping $a_1$ and $a_2$. The contribution of each graph to the path integral \eqref{rg} can be derived from the volumes mentioned above, which are:
\begin{align}
&\mathcal{Z}_{1}^{\text{(0,5)}}(\beta,a_1)=\frac{1}{48} \beta  \left(\frac{2 \sqrt{\beta }  \left(a_{1}^2+4 \beta
   \right)}{\sqrt{\pi }}\exp\left(-\frac{a_{1}^2}{4 \beta }\right)-a_1 \left(a_{1}^2+6 \beta \right) \text{erfc}\left(\frac{a_1}{2 \sqrt{\beta }}\right)\right),\nonumber
\\&\mathcal{Z}_{1}^{\text{(0,4)}}(\beta,a_1)=\frac{1}{16} \beta a_1 \left(\left(a_{1}^2+2 \beta \right) \text{erfc}\left(\frac{a_1}{2 \sqrt{\beta
   }}\right)-\frac{2 \sqrt{\beta } a_1 }{\sqrt{\pi }}\exp\left(-\frac{a_{1}^2}{4 \beta }\right)\right),\nonumber
\\
&\mathcal{Z}_{1}^{\text{(0,3)}}(\beta,a_1)=\frac{\beta ^{3/2} a_{1}^2  }{6\sqrt{\pi }}\exp\left(-\frac{a_{1}^2}{4 \beta }\right)-\frac{1}{12} \beta  a_{1}^3 
   \text{erfc}\left(\frac{a_{1}}{2 \sqrt{\beta }}\right),\nonumber
\\&\mathcal{Z}_{1}^{\text{(0,2)}}(\beta,a_1)=\frac{1}{24} \beta  a_{1}^3 \text{erfc}\left(\frac{a_{1}}{2 \sqrt{\beta }}\right),\nonumber\\
&\mathcal{Z}_{1}^{\text{(1,1)}}(\beta,a_1)=\frac{\beta ^{3/2}  }{6
   \sqrt{\pi }}\left(4 \beta -\exp\left(-\frac{a_{1}^2}{4 \beta }\right)\left(a_{1}^2+4 \beta \right)\right).
\end{align}
The above expressions reveal that  Kontsevich graphs which contain the term $\exp\left(-\frac{a_{1}^2}{4 \beta }\right)$ relate the black hole's age to the baby universe's length. Explicitly, the path integral $\mathcal{Z}_{1}^{\text{(0,2)}}(\beta, a_1)$ does not exhibit the feature of baby universe emission. One can also see that other graphs whose volume can be obtained by swapping  $a_1$ and $a_2$ in \eqref{hr}, exhibit this feature.

\section{Concluding remarks}\label{con}

In this paper, Jackiw-Teitelboim gravity path integrals were explored, focusing particularly on the tunneling process between aging and younger black holes through the emission of baby universes. The main motivation for studying this subject was to better understand the relationship between the effective age of a black hole and the geometrical structures formed due to these tunneling processes. By examining the path integrals for geometries with more complex structures, including those with higher-genus structures and multiple geodesic boundaries, an attempt was made to provide a clear explanation of the physical implications of these quantum gravity effects.

The main observation is that while the path integral of the trumpet geometry in Jackiw-Teitelboim gravity can be interpreted as a tunneling amplitude, this interpretation becomes less straightforward for geometries with multiple baby universes and nontrivial Weil-Petersson volumes. Indeed, by making use of Kontsevich’s approach, which utilizes ribbon graphs to compute Weil-Petersson volumes in the Airy limit, specific regions of the moduli space that meaningfully contribute to the tunneling process were identified. This approach allowed for the extraction of transition amplitudes that directly relate the emission of baby universes to the effective age of black holes.

In fact, it has been demonstrated that not all contributions to the Weil-Petersson volume correspond to a physically meaningful tunneling process. Only certain Kontsevich graphs exhibited the expected relationship between black hole age and baby universe emission. This finding highlights the need to restrict moduli space considerations to specific configurations that contribute to observable physical phenomena. This restriction is in agreement with the study by Stanford and Yang \cite{Stanford:2022fdt}, which introduced a delta function into the path integral of the three-holed sphere to focus on specific regions of the moduli space associated with firewall geometries (see also \cite{Zolfi:2024ldx} for the case of a five-holed sphere). This delta function isolates geometries containing wormholes of specific lengths to calculate the probability of encountering a firewall.

These results help to better understand how black holes change over time. The creation of baby universes makes black holes younger and offers a simpler way to approach the black hole information problem and the nature of space-time connectivity. Moreover, these results may suggest that this procedure is highly sensitive to the underlying geometric constraints, implying that not all configurations contribute equally to observable quantum processes.
\section*{Acknowledgments}
We thank Seyed Amirhossein Mousavi, Reza Pirmoradiyan, Mohammad Reza Tanhaei, and Mohammad Javad Vasli for their valuable support and insightful discussions. The work of M. A. is supported by the Iran National Science Foundation (INSF) under project No. 4023620.
Additionally, OpenAI's ChatGPT was used to enhance the clarity of this manuscript.
\appendix	
\section{Kontsevich graphs of four-holed sphere}\label{iii}
To derive the WP volume given in equation \eqref{volf} at the Airy limit, one should consider Kontsevich graphs with six edges and four vertices. The geodesic boundaries $a_1, a_2, a_3$ and $a_4$ are indicated in the ribbon graphs by curves colored green, orange, blue, and red, respectively.
In the simplest form of a ribbon graph,  geodesic boundaries $a_1, a_2$ and $a_3$,  are symmetric in their roles, while the geodesic boundary $a_4$ is bigger than the sum $a_1+a_2+a_3$. The  Laplace transform of the WP volume of this  graph  is:
\vspace{1cm}

\begin{figure}[H]
	\centering
	\begin{overpic}
		[width=0.42\textwidth]{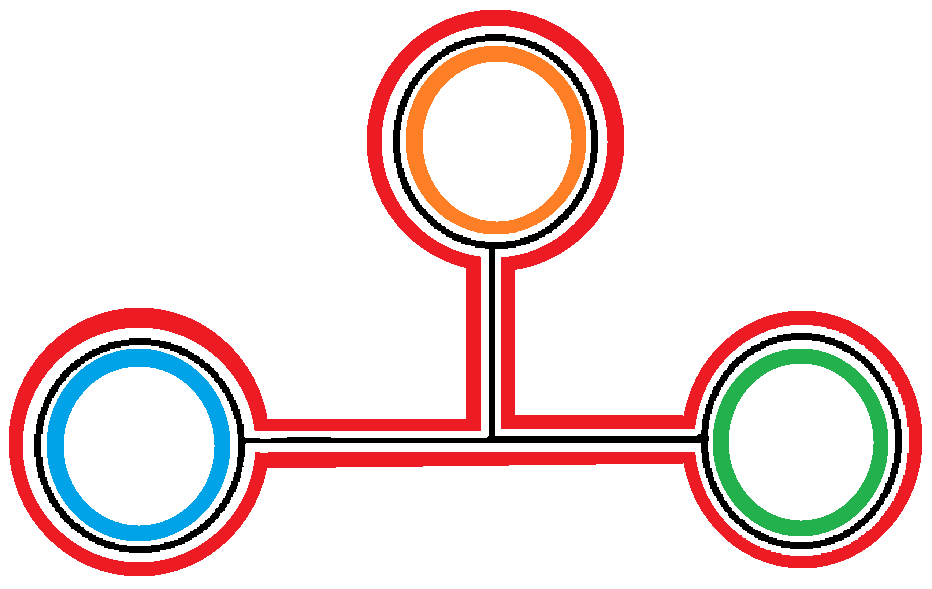}
		\put (99,17) {$\displaystyle \frac{1}{z_1+z_4}$}	
		\put (61,3) {$\displaystyle \frac{1}{2z_4}$}	
		\put (41,24) {$\displaystyle \frac{1}{2z_4}$}	
		\put (37,3) {$\displaystyle \frac{1}{2z_4}$}
		\put (42,67) {$\displaystyle \frac{1}{z_2+z_4}$}
		\put (-18,15) {$\displaystyle \frac{1}{z_3+z_{4}}$}		\put (-41,15) {$\displaystyle \tilde{	V}^{\text{I}}_{0,4}(\textbf{z})=	$}			
	\end{overpic}
	\caption{In Kontsevich graph I, the geodesic boundaries $a_1$, $a_2$, and $a_3$ display similar behavior, while $a_4$ behaves differently.}	\label{ocp}
\end{figure}

Following the application of the inverse Laplace transform, the result is:
\begin{align}\label{lapry}
	V_{0,4}^{\text{I}}(\textbf{a})&=\int_{\gamma+\text{i}\mathbb{R}}\prod_{i=1}^{4}\frac{\text{d}z_i}{2\pi \text{i}}e^{a_iz_i}	\frac{c_{1}}{z_4^{3}\left(z_1+z_4 \right)\left(z_2+z_4 \right)\left(z_3+z_4 \right)   }\nonumber\\
	&=\frac{c_1}{2} \left(a_1+a_2+a_3-a_4\right){}^2 \theta \left(a_4-a_1-a_2-a_3\right).
\end{align}
The coefficients $c_i $ (for $i=1,\dots, 5$) are determined by the symmetry factor of each ribbon graph. As expected from figure \ref{ocp} and the previous result, it is clear that  $a_1, a_2$ and $a_3$ display similar behavior, while $a_4$ behaves differently in $V_{0,4}^{\text{I}}$. Due to the presence of the theta function in $V_{0,4}^{\text{I}}$, we will focus on configurations of ribbon graphs that produce this theta function to simplify the derivation of \eqref{volf} in the Airy limit.

\begin{figure}[H]
	\centering
	\begin{overpic}
		[width=0.42\textwidth]{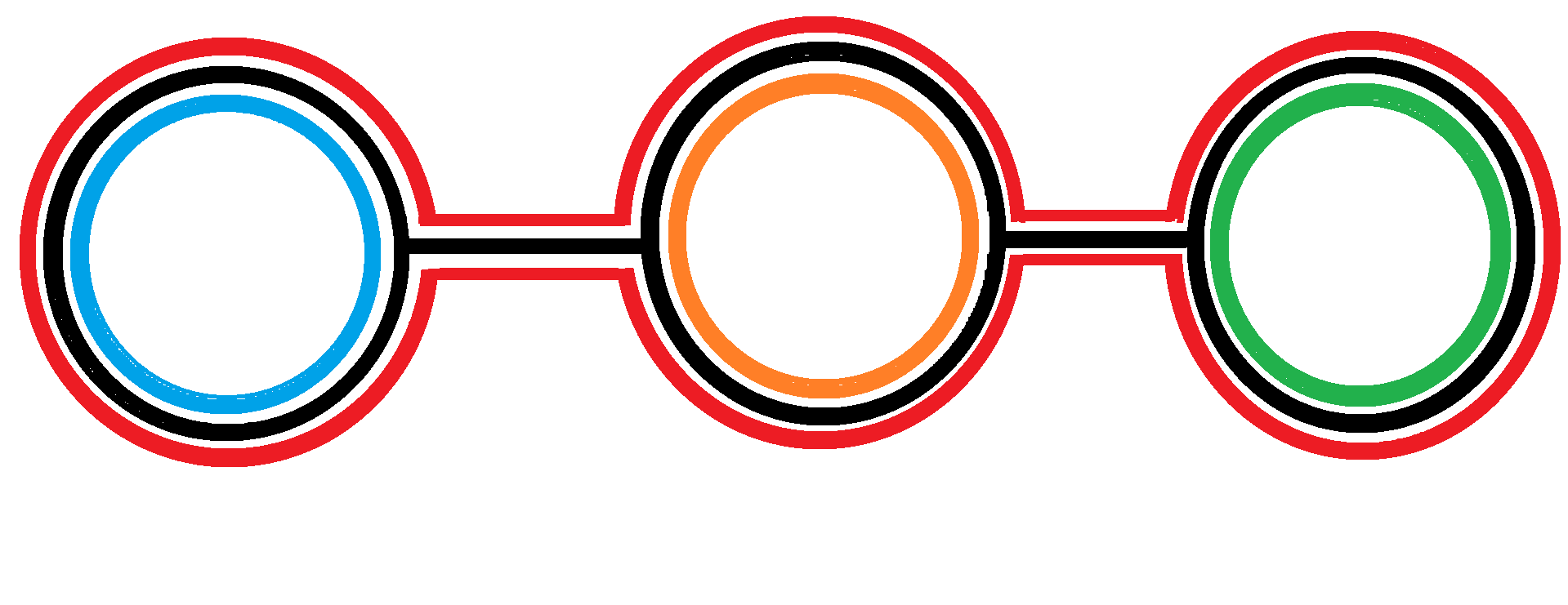}
		\put (103,16) {$\displaystyle \frac{1}{z_1+z_4}$}	
		\put (66,12) {$\displaystyle \frac{1}{2z_4}$}	
		\put (43,43) {$\displaystyle \frac{1}{z_2+z_4}$}	
		\put (43,2) {$\displaystyle \frac{1}{z_2+z_4}$}
		\put (30,12) {$\displaystyle \frac{1}{2z_4}$}
		\put (-19,16) {$\displaystyle \frac{1}{z_3+z_4}$}		\put (-41,15) {$\displaystyle V^{\text{II}}_{0,4}(\textbf{z})=	$}	
	\end{overpic}
	\\
\caption{In Kontsevich graph II, the geodesic boundaries $a_1$ and $a_2$ display similar behavior, while   $a_3$ and $a_4$ behaves differently.}
	\label{GG}
\end{figure}

To construct a new Kontsevich graph, starting from graph I, one can first collapse the vertical edge, merging the two vertices into a single vertex. Subsequently, one obtains ribbon graph II by expanding this new vertex along the horizontal direction (see figure \ref{GG}). In this ribbon graph, the contribution of the geodesic boundary $a_2$ in $V^{\text{II}}_{0,4}(\textbf{z})$ differs from that of $a_1$ and $a_3$ and the power of propagator $1/z_4$ decreases by one order.
\vspace{1cm}

\begin{figure}[H]
	\centering
	\begin{overpic}
		[width=0.45\textwidth]{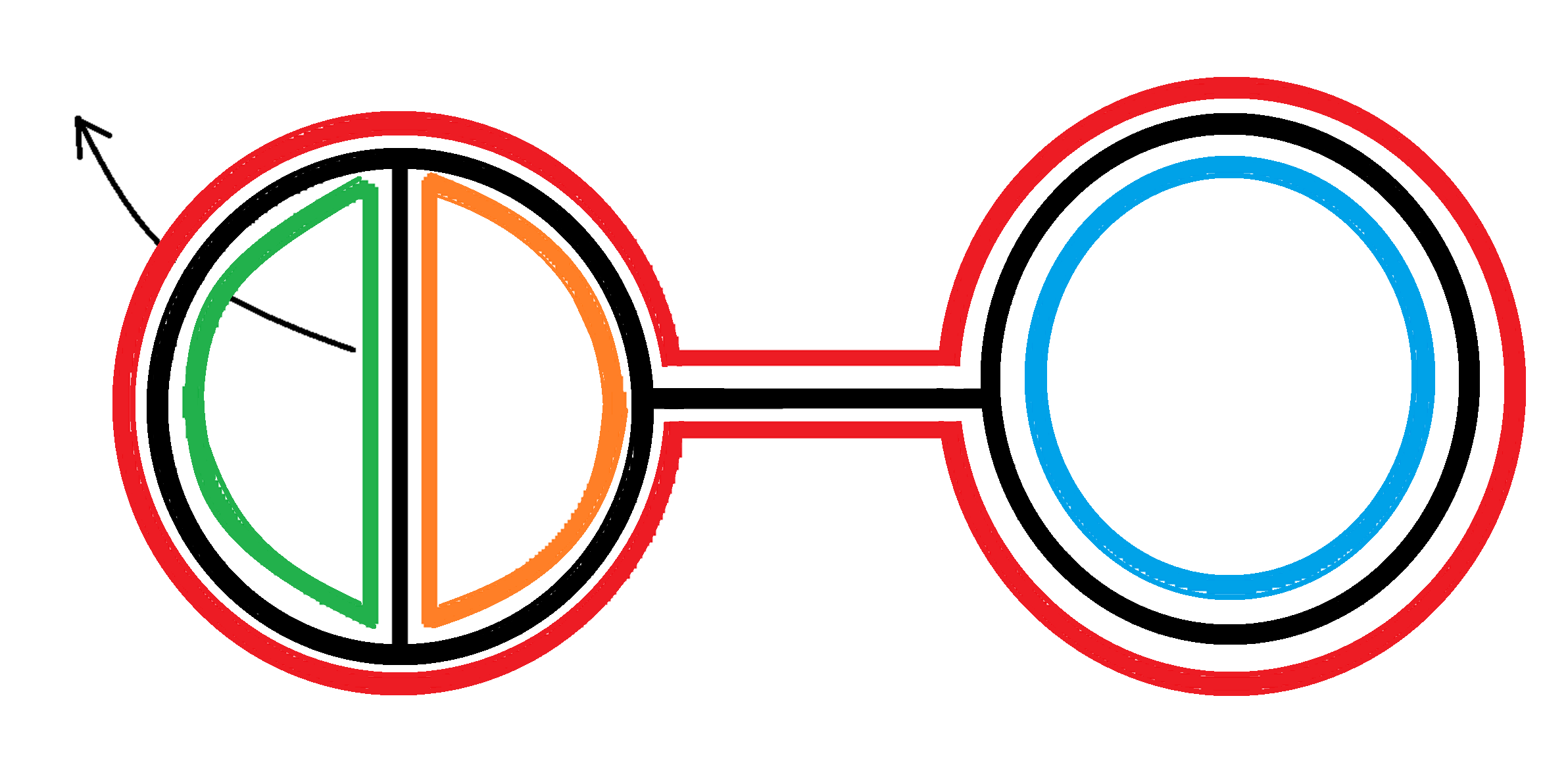}
		\put (101,17) {$\displaystyle \frac{1}{z_3+z_4}$}	
		\put (-13,15) {$\displaystyle \frac{1}{z_1+z_4}$}	
		\put (-13,40) {$\displaystyle \frac{1}{z_1+z_2}$}	
		\put (34,1) {$\displaystyle \frac{1}{z_2+z_4}$}
		\put (27,46) {$\displaystyle \frac{1}{z_2+z_4}$}
		\put (47,31) {$\displaystyle \frac{1}{2z_4}$}		\put (-41,15) {$\displaystyle \tilde{	V}^{\text{IIIA}}_{0,4}(\textbf{z})=	$}		
	\end{overpic}
\caption{Kontsevich graph IIIA}	
	\label{loop}
\end{figure}

The contribution of ribbon  graph II to the volume is:
\begin{align}\label{ldry}
	V_{0,4}^{\text{II}}(\textbf{a})&=	\int_{\gamma+\text{i}\mathbb{R}}\prod_{i=1}^{4}\frac{\text{d}z_i}{2\pi \text{i}}e^{a_iz_i}\frac{c_2}{z_4^{2}\left( z_1+z_4\right)\left(z_2+z_4 \right)^{2}\left(z_3+z_4 \right)   }\nonumber\\
	&=c_2a_2 \left(a_4-a_1-a_2-a_3\right) \theta \left(a_4-a_1-a_2-a_3\right).
\end{align}	
If $a_2$ is replaced with $a_1$ and $a_3$, the total contribution of this configuration to the volume is:
\begin{align}\label{dfy2}
	V_{0,4}^{\text{ II}_{\text{total}}}(\textbf{a})=	c_2\left( a_1+a_2+a_3\right)  \left(a_4-a_1-a_2-a_3\right) \theta \left(a_4-a_1-a_2-a_3\right).
\end{align}	
The Whitehead method is now applied again to obtain another ribbon graph. By contracting the left horizontal edge of ribbon graph II and expanding vertically, ribbon graph IIIA is obtained (see figure \ref{loop}).
\\
 In this  graph, the power of $1/z_4$ is one and its contribution to the volume of moduli space is:
\begin{align}\label{lrt}
	V_{0,4}^{\text{IIIA}}(\textbf{a})&=\int_{\gamma+\text{i}\mathbb{R}}\prod_{i=1}^{4}\frac{\text{d}z_i}{2\pi \text{i}}e^{a_iz_i}\frac{c_3}{z_{4}\left( z_1+z_4\right)\left( z_1+z_2\right)\left( z_2+z_4\right)^{2} \left( z_3+z_4\right)}\nonumber\\
	&=\frac{c_3}{8}\bigg\lbrace\left(a_1-a_2+a_3-a_4\right){}^2 \theta \left(a_1-a_2,a_4-a_1+a_2-a_3\right)
	\nonumber\\
	&\qquad\quad+4 a_2 \left(a_1+a_2+a_3-a_4\right) \theta
	\left(a_4-a_1-a_2-a_3\right)
	\nonumber\\
	&\qquad\quad-\left(a_1+a_2+a_3-a_4\right){}^2 \theta
	\left(a_4-a_1-a_2-a_3\right)
	\nonumber\\
	&\qquad\quad
	-\left(a_1-a_2-a_3+a_4\right){}^2 \theta
	\left(a_1-a_2,a_1-a_2-a_3+a_4\right)\nonumber\\
	&\qquad\quad+4 a_2 \left(-a_1+a_2+a_3-a_4\right) \theta \left(a_1-a_2,a_1-a_2-a_3+a_4\right)\nonumber\\
	&\qquad\quad+4 a_1
	\left(a_1-a_2-a_3+a_4\right) \theta \left(a_1-a_2,a_1-a_2-a_3+a_4\right)\nonumber\\
	&\qquad\quad+\left(a_1-a_2-a_3+a_4\right){}^2 \theta \left(a_1-a_2-a_3+a_4\right)\nonumber\\
	&\qquad\quad+4 a_1
	\left(-a_1+a_2+a_3-a_4\right) \theta \left(a_1-a_2-a_3+a_4\right)\nonumber\\
	&\qquad\quad+4 a_2 \left(a_1-a_2-a_3+a_4\right) \theta \left(a_1-a_2-a_3+a_4\right)\bigg\rbrace.
\end{align}	
Assuming $a_4 > a_3 + a_2 + a_1$, the relation above simplifies to:
\begin{align}\label{iiia}
	V_{0,4}^{\text{IIIA}}(\textbf{a})=\frac{c_3}{2}\bigg\lbrace a_2^2 \theta \left(a_1-a_2\right)- a_1 \left(a_1-2 a_2\right) \theta \left(a_2-a_1\right)\bigg\rbrace,
\end{align}
and other permutations among the geodesic boundaries yield:

\begin{align}\label{c3}
	c_3\bigg\lbrace a_1a_2+a_1a_3+a_2a_3\bigg\rbrace.
\end{align}

\begin{figure}[H]
	\centering
	\begin{overpic}
		[width=0.53\textwidth]{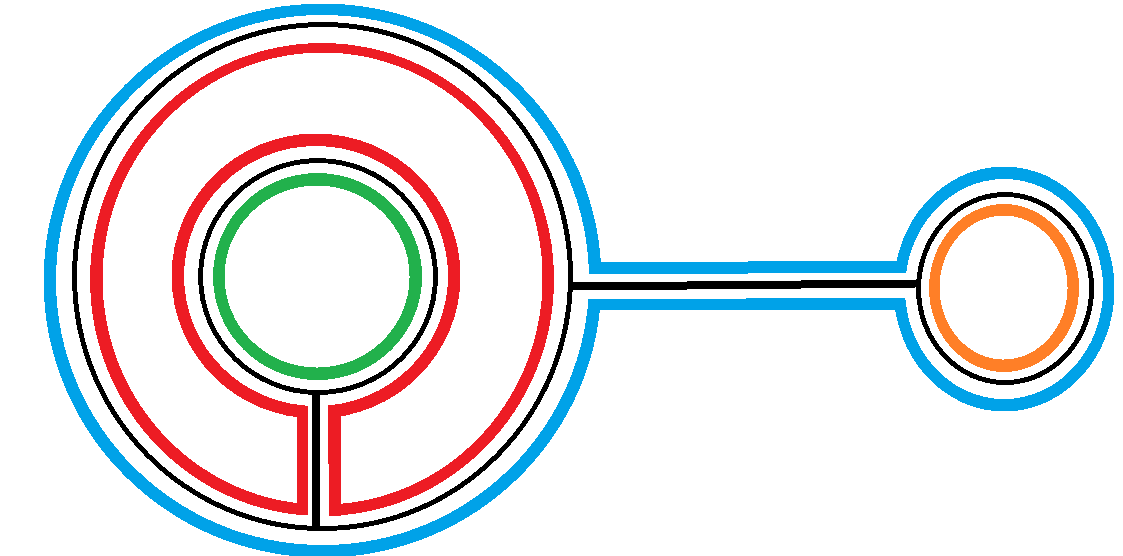}
		\put (100,22) {$\displaystyle \frac{1}{z_2+z_3}$}	
		\put (-15,22) {$\displaystyle \frac{1}{z_3+z_4}$}	
		\put (32,8) {$\displaystyle z_4^{-1}$}	
		\put (44,3) {$\displaystyle \frac{1}{z_3+z_4}$}
		\put (17,38) {$\displaystyle \left( z_1+z_4\right)^{-1} $}
		\put (63,32) {$\displaystyle \frac{1}{2z_3}$}	
		\put (-38,22) {$\displaystyle \tilde{	V}^{\text{IIIB}}_{0,4}(\textbf{z})=	$}		
	\end{overpic}
	\caption{Kontsevich graph IIIB}
	\label{liopr}
\end{figure}

\begin{figure}[H]
	\centering
	\begin{overpic}
		[width=0.45\textwidth]{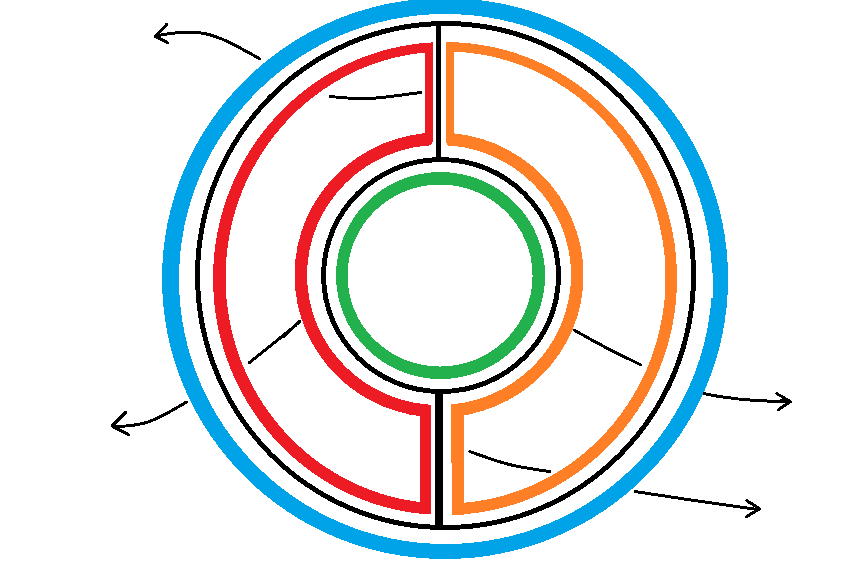}
		\put (86,42) {$\displaystyle \frac{1}{z_2+z_3}$}	
		\put (-2,42) {$\displaystyle \frac{1}{z_3+z_4}$}	
		\put (90,4) {$\displaystyle  \frac{1}{z_2+z_4}  $}	
		\put (-6,15) {$\displaystyle \frac{1}{z_1+z_4}$}
		\put (-5,62) {$\displaystyle \frac{1}{z_2+z_4} $}
		\put (93,20) {$\displaystyle \frac{1}{z_1+z_2}$}	\put (-36,15) {$\displaystyle \tilde{	V}^{\text{IV}}_{0,4}(\textbf{z})=	$}		
	\end{overpic}
	\caption{Kontsevich graph IV}
	\label{liop}
\end{figure}

\begin{figure}[h]
	\centering
	\begin{overpic}
		[width=0.45\textwidth]{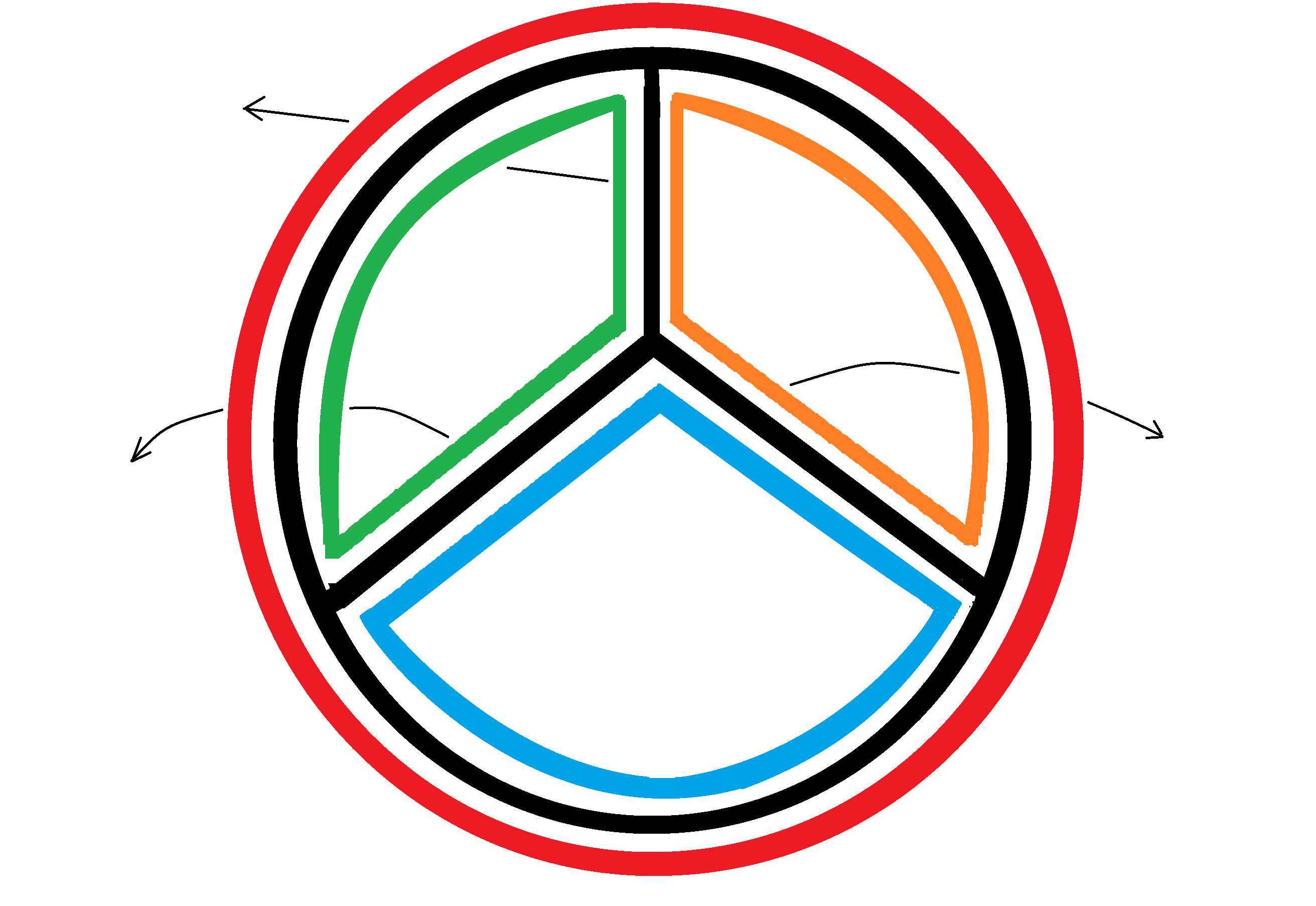}
		\put (80,59) {$\displaystyle \frac{1}{z_2+z_4}$}	
		\put (-1,66) {$\displaystyle \frac{1}{z_1+z_2}$}	
		\put (85,32) {$\displaystyle \frac{1}{z_2+z_3} $}	
		\put (64,3) {$\displaystyle \frac{1}{z_3+z_4}$}
		\put (0,27) {$\displaystyle \frac{1}{z_1+z_3}$}
		\put (0,50) {$\displaystyle \frac{1}{z_1+z_4}$}	
		\put (-28,27) {$\displaystyle \tilde{	V}^{\text{V}}_{0,4}(\textbf{z})=	$}	
	\end{overpic}
	\caption{Kontsevich graph V}
	\label{lofp}
\end{figure}

Another configuration, similar to graph IIIA, also includes the propagator $1/z_4$ (see figure \ref{liopr}). However, this configuration differs from graph IIIA due to variations in the orientations of the geodesic boundaries. For clarification, consider expression \eqref{lapry} for graph I, where $1/z_4^3$ ensures that $a_4 > a_3$, $a_4 > a_2$, and $a_4 > a_1$. In graph II, $1/z_4^2$ appears, while in graph IIIA, expression \eqref{lrt} contains $1/z_4$. The presence of $1/z_4^n$ ($n = 1, 2, 3$) in these expressions is crucial. For simplicity, the analysis of the ribbon graphs is restricted to the condition $\theta(a_4 - a_1 - a_2 - a_3)$, where ribbon graphs containing propagators $1/z_4^n$ contribute to this condition. In ribbon graphs IV and V (see figures \ref{liop} and \ref{lofp}), no $1/z_4^n$ terms are present, and their contributions are zero. Interestingly, graph IIIB contributes similarly to IIIA, as both contain $1/z_4$, in contrast to graph II, which contains $1/z_4^2$.
The WP volume of this  graph is: 
\begin{align}\label{lrk}
	V_{0,4}^{\text{IIIB}}(\textbf{a})&=\int_{\gamma+\text{i}\mathbb{R}}\prod_{i=1}^{4}\frac{\text{d}z_i}{2\pi \text{i}}e^{a_iz_i}\frac{1}{z_3 \left(z_2+z_3\right) z_4 \left(z_1+z_4\right) \left(z_3+z_4\right){}^2}\nonumber\\
	&=\frac{\tilde{c}_{3}}{2} \theta \left(a_3-a_2\right) \bigg(\left(a_1-a_4\right){}^2 \theta
	\left(a_4-a_1\right)-\left(a_1+a_2-a_3-a_4\right)\nonumber\\
	&\qquad\qquad\qquad\qquad \times\left(a_1-a_2+a_3-a_4\right) \theta \left(a_4-a_1+a_2-a_3\right)\bigg).
\end{align}	
For the case $a_4>a_3+a_2+a_1$ and $a_3>a_2>a_1$ the above expression simplifies to:
\begin{align}\label{iiib}
	V_{0,4}^{\text{IIIB}}(\textbf{a})&=\frac{\tilde{c}_{3}}{2}\left( a_2^2-2a_2 a_3+a_3^2\right).
\end{align}	
Considering all permutations gives:
\begin{align}\label{c4}
	\tilde{c}_{3}\bigg\lbrace a_1^2+a_2^2+a_3^2-a_1a_2-a_1a_3 -a_2 a_3\bigg\rbrace.
\end{align}
By setting $c_3=2\tilde{c}_{3}$, and using equations \eqref{c3} and \eqref{c4}, we obtain:
\begin{align}\label{w}
	V_{0,4}^{\text{III}_{\text{total}}}(\textbf{a})	=\frac{c_{3}}{2}\bigg\lbrace a_1^{2}+a_2^{2}+a_3^{2}+a_1a_2+a_1a_3+a_2a_3\bigg\rbrace.
\end{align}
We can also consider two additional Kontsevich graphs that do not include the $1/z_4$ factor. Hence, the contribution of these graphs to the volume of the moduli space is zero by the assumption $a_4>a_3+a_2+a_1$ (see figures \ref{liop} and \ref{lofp}).
By summing the contributions of all volumes in \eqref{lapry}, \eqref{dfy2}, and \eqref{w}, and with $c_1$ set equal to $c_2$ and $c_3$ set to 2, the following result is obtained:
\begin{align}
	V_{0,4}^{\text{Airy}}=\frac{1}{2}	\left(a_{1}^2+a_{2}^{2}+a_{3}^{2}+a_{4}^{2}\right)  \theta \left(a_4-a_1-a_2-a_3\right).
\end{align} 
\bibliographystyle{
utphys.bst}
\bibliography{Ref}

\providecommand{\href}[2]{#2}\begingroup\raggedright\begin{thebibliography}{10}

\bibitem{haw}
S.~Hawking, ``Quantum coherence down the wormhole,'' \href{https://dx.doi.org/https://doi.org/10.1016/0370-2693(87)90028-1}{{\em Physics Letters B} {\bfseries 195} no.~3, (1987) 337--343}. \url{https://www.sciencedirect.com/science/article/pii/0370269387900281}.

\bibitem{Lavrelashvili:1987jg}
G.~V. Lavrelashvili, V.~A. Rubakov, and P.~G. Tinyakov, ``{Disruption of Quantum Coherence upon a Change in Spatial Topology in Quantum Gravity},'' {\em JETP Lett.} {\bfseries 46} (1987) 167--169.

\bibitem{Maldacena:2004rf}
J.~M. Maldacena and L.~Maoz, ``{Wormholes in AdS},'' \href{https://dx.doi.org/10.1088/1126-6708/2004/02/053}{{\em JHEP} {\bfseries 02} (2004) 053}, \href{https://arxiv.org/abs/hep-th/0401024}{{\ttfamily arXiv:hep-th/0401024}}.

\bibitem{Arkani-Hamed:2007cpn}
N.~Arkani-Hamed, J.~Orgera, and J.~Polchinski, ``{Euclidean wormholes in string theory},'' \href{https://dx.doi.org/10.1088/1126-6708/2007/12/018}{{\em JHEP} {\bfseries 12} (2007) 018}, \href{https://arxiv.org/abs/0705.2768}{{\ttfamily arXiv:0705.2768 [hep-th]}}.

\bibitem{Saad:2018bqo}
P.~Saad, S.~H. Shenker, and D.~Stanford, ``{A semiclassical ramp in SYK and in gravity},'' \href{https://arxiv.org/abs/1806.06840}{{\ttfamily arXiv:1806.06840 [hep-th]}}.

\bibitem{Mathur:2009hf}
S.~D. Mathur, ``{The Information paradox: A Pedagogical introduction},'' \href{https://dx.doi.org/10.1088/0264-9381/26/22/224001}{{\em Class. Quant. Grav.} {\bfseries 26} (2009) 224001}, \href{https://arxiv.org/abs/0909.1038}{{\ttfamily arXiv:0909.1038 [hep-th]}}.

\bibitem{Almheiri:2012rt}
A.~Almheiri, D.~Marolf, J.~Polchinski, and J.~Sully, ``{Black Holes: Complementarity or Firewalls?},'' \href{https://dx.doi.org/10.1007/JHEP02(2013)062}{{\em JHEP} {\bfseries 02} (2013) 062}, \href{https://arxiv.org/abs/1207.3123}{{\ttfamily arXiv:1207.3123 [hep-th]}}.

\bibitem{Bousso:2012as}
R.~Bousso, ``{Complementarity Is Not Enough},'' \href{https://dx.doi.org/10.1103/PhysRevD.87.124023}{{\em Phys. Rev. D} {\bfseries 87} no.~12, (2013) 124023}, \href{https://arxiv.org/abs/1207.5192}{{\ttfamily arXiv:1207.5192 [hep-th]}}.

\bibitem{Nomura:2012sw}
Y.~Nomura, J.~Varela, and S.~J. Weinberg, ``{Complementarity Endures: No Firewall for an Infalling Observer},'' \href{https://dx.doi.org/10.1007/JHEP03(2013)059}{{\em JHEP} {\bfseries 03} (2013) 059}, \href{https://arxiv.org/abs/1207.6626}{{\ttfamily arXiv:1207.6626 [hep-th]}}.

\bibitem{Verlinde:2012cy}
E.~Verlinde and H.~Verlinde, ``{Black Hole Entanglement and Quantum Error Correction},'' \href{https://dx.doi.org/10.1007/JHEP10(2013)107}{{\em JHEP} {\bfseries 10} (2013) 107}, \href{https://arxiv.org/abs/1211.6913}{{\ttfamily arXiv:1211.6913 [hep-th]}}.

\bibitem{Papadodimas:2012aq}
K.~Papadodimas and S.~Raju, ``{An Infalling Observer in AdS/CFT},'' \href{https://dx.doi.org/10.1007/JHEP10(2013)212}{{\em JHEP} {\bfseries 10} (2013) 212}, \href{https://arxiv.org/abs/1211.6767}{{\ttfamily arXiv:1211.6767 [hep-th]}}.

\bibitem{JACKIW1985343}
R.~Jackiw, ``Lower dimensional gravity,'' \href{https://dx.doi.org/https://doi.org/10.1016/0550-3213(85)90448-1}{{\em Nuclear Physics B} {\bfseries 252} (1985) 343--356}. \url{https://www.sciencedirect.com/science/article/pii/0550321385904481}.

\bibitem{Teitelboim:1983ux}
C.~Teitelboim, ``{Gravitation and Hamiltonian Structure in Two Space-Time Dimensions},'' \href{https://dx.doi.org/10.1016/0370-2693(83)90012-6}{{\em Phys. Lett. B} {\bfseries 126} (1983) 41--45}.

\bibitem{Nojiri:2024ycf}
S.~Nojiri and S.~D. Odintsov, ``{Non-metricity approach to Jackiw-Teitelboim gravity},'' \href{https://arxiv.org/abs/2410.20747}{{\ttfamily arXiv:2410.20747 [hep-th]}}.

\bibitem{yang}
Z.~Yang, ``{The Quantum Gravity Dynamics of Near Extremal Black Holes},'' \href{https://dx.doi.org/10.1007/JHEP05(2019)205}{{\em JHEP} {\bfseries 05} (2019) 205}, \href{https://arxiv.org/abs/1809.08647}{{\ttfamily arXiv:1809.08647 [hep-th]}}.

\bibitem{kit}
A.~Kitaev and S.~J. Suh, ``{Statistical mechanics of a two-dimensional black hole},'' \href{https://dx.doi.org/10.1007/JHEP05(2019)198}{{\em JHEP} {\bfseries 05} (2019) 198}, \href{https://arxiv.org/abs/1808.07032}{{\ttfamily arXiv:1808.07032 [hep-th]}}.

\bibitem{Bagrets:2017pwq}
D.~Bagrets, A.~Altland, and A.~Kamenev, ``{Power-law out of time order correlation functions in the SYK model},'' \href{https://dx.doi.org/10.1016/j.nuclphysb.2017.06.012}{{\em Nucl. Phys. B} {\bfseries 921} (2017) 727--752}, \href{https://arxiv.org/abs/1702.08902}{{\ttfamily arXiv:1702.08902 [cond-mat.str-el]}}.

\bibitem{har}
D.~Harlow and D.~Jafferis, ``{The Factorization Problem in Jackiw-Teitelboim Gravity},'' \href{https://dx.doi.org/10.1007/JHEP02(2020)177}{{\em JHEP} {\bfseries 02} (2020) 177}, \href{https://arxiv.org/abs/1804.01081}{{\ttfamily arXiv:1804.01081 [hep-th]}}.

\bibitem{Saad:2019lba}
P.~Saad, S.~H. Shenker, and D.~Stanford, ``{JT gravity as a matrix integral},'' \href{https://arxiv.org/abs/1903.11115}{{\ttfamily arXiv:1903.11115 [hep-th]}}.

\bibitem{Stanford:2019vob}
D.~Stanford and E.~Witten, ``{JT gravity and the ensembles of random matrix theory},'' \href{https://dx.doi.org/10.4310/ATMP.2020.v24.n6.a4}{{\em Adv. Theor. Math. Phys.} {\bfseries 24} no.~6, (2020) 1475--1680}, \href{https://arxiv.org/abs/1907.03363}{{\ttfamily arXiv:1907.03363 [hep-th]}}.

\bibitem{Witten:2020wvy}
E.~Witten, ``{Matrix Models and Deformations of JT Gravity},'' \href{https://dx.doi.org/10.1098/rspa.2020.0582}{{\em Proc. Roy. Soc. Lond. A} {\bfseries 476} no.~2244, (2020) 20200582}, \href{https://arxiv.org/abs/2006.13414}{{\ttfamily arXiv:2006.13414 [hep-th]}}.

\bibitem{Stanford:2022fdt}
D.~Stanford and Z.~Yang, ``{Firewalls from wormholes},'' \href{https://arxiv.org/abs/2208.01625}{{\ttfamily arXiv:2208.01625 [hep-th]}}.

\bibitem{Saad:2019pqd}
P.~Saad, ``{Late Time Correlation Functions, Baby Universes, and ETH in JT Gravity},'' \href{https://arxiv.org/abs/1910.10311}{{\ttfamily arXiv:1910.10311 [hep-th]}}.

\bibitem{Giddings:1987cg}
S.~B. Giddings and A.~Strominger, ``{Axion Induced Topology Change in Quantum Gravity and String Theory},'' \href{https://dx.doi.org/10.1016/0550-3213(88)90446-4}{{\em Nucl. Phys. B} {\bfseries 306} (1988) 890--907}.

\bibitem{COLEMAN1988867}
S.~Coleman, ``Black holes as red herrings: Topological fluctuations and the loss of quantum coherence,'' \href{https://dx.doi.org/https://doi.org/10.1016/0550-3213(88)90110-1}{{\em Nuclear Physics B} {\bfseries 307} no.~4, (1988) 867--882}. \url{https://www.sciencedirect.com/science/article/pii/0550321388901101}.

\bibitem{Susskind:2015toa}
L.~Susskind, ``{The Typical-State Paradox: Diagnosing Horizons with Complexity},'' \href{https://dx.doi.org/10.1002/prop.201500091}{{\em Fortsch. Phys.} {\bfseries 64} (2016) 84--91}, \href{https://arxiv.org/abs/1507.02287}{{\ttfamily arXiv:1507.02287 [hep-th]}}.

\bibitem{Zolfi:2024ldx}
H.~Zolfi, ``{Firewalls from wormholes in higher genus},'' \href{https://dx.doi.org/10.1007/JHEP05(2024)039}{{\em JHEP} {\bfseries 05} (2024) 039}, \href{https://arxiv.org/abs/2401.04476}{{\ttfamily arXiv:2401.04476 [hep-th]}}.

\bibitem{Hartman:2013qma}
T.~Hartman and J.~Maldacena, ``{Time Evolution of Entanglement Entropy from Black Hole Interiors},'' \href{https://dx.doi.org/10.1007/JHEP05(2013)014}{{\em JHEP} {\bfseries 05} (2013) 014}, \href{https://arxiv.org/abs/1303.1080}{{\ttfamily arXiv:1303.1080 [hep-th]}}.

\bibitem{Susskind:2014rva}
L.~Susskind, ``{Computational Complexity and Black Hole Horizons},'' \href{https://dx.doi.org/10.1002/prop.201500092}{{\em Fortsch. Phys.} {\bfseries 64} (2016) 24--43}, \href{https://arxiv.org/abs/1403.5695}{{\ttfamily arXiv:1403.5695 [hep-th]}}. [Addendum: Fortsch.Phys. 64, 44--48 (2016)].

\bibitem{Iliesiu:2021ari}
L.~V. Iliesiu, M.~Mezei, and G.~S\'arosi, ``{The volume of the black hole interior at late times},'' \href{https://dx.doi.org/10.1007/JHEP07(2022)073}{{\em JHEP} {\bfseries 07} (2022) 073}, \href{https://arxiv.org/abs/2107.06286}{{\ttfamily arXiv:2107.06286 [hep-th]}}.

\bibitem{Alishahiha:2022kzc}
M.~Alishahiha, S.~Banerjee, and J.~Kames-King, ``{Complexity via replica trick},'' \href{https://dx.doi.org/10.1007/JHEP08(2022)181}{{\em JHEP} {\bfseries 08} (2022) 181}, \href{https://arxiv.org/abs/2205.01150}{{\ttfamily arXiv:2205.01150 [hep-th]}}.

\bibitem{Alishahiha:2022exn}
M.~Alishahiha and S.~Banerjee, ``{On the saturation of late-time growth of complexity in supersymmetric JT gravity},'' \href{https://dx.doi.org/10.1007/JHEP01(2023)134}{{\em JHEP} {\bfseries 01} (2023) 134}, \href{https://arxiv.org/abs/2209.02441}{{\ttfamily arXiv:2209.02441 [hep-th]}}.

\bibitem{Iliesiu:2024cnh}
L.~V. Iliesiu, A.~Levine, H.~W. Lin, H.~Maxfield, and M.~Mezei, ``{On the non-perturbative bulk Hilbert space of JT gravity},'' \href{https://arxiv.org/abs/2403.08696}{{\ttfamily arXiv:2403.08696 [hep-th]}}.

\bibitem{Belin:2021bga}
A.~Belin, R.~C. Myers, S.-M. Ruan, G.~S\'arosi, and A.~J. Speranza, ``{Does Complexity Equal Anything?},'' \href{https://dx.doi.org/10.1103/PhysRevLett.128.081602}{{\em Phys. Rev. Lett.} {\bfseries 128} no.~8, (2022) 081602}, \href{https://arxiv.org/abs/2111.02429}{{\ttfamily arXiv:2111.02429 [hep-th]}}.

\bibitem{Blommaert:2024ftn}
A.~Blommaert, C.-H. Chen, and Y.~Nomura, ``{Firewalls at exponentially late times},'' \href{https://dx.doi.org/10.1007/JHEP10(2024)131}{{\em JHEP} {\bfseries 10} (2024) 131}, \href{https://arxiv.org/abs/2403.07049}{{\ttfamily arXiv:2403.07049 [hep-th]}}.

\bibitem{Bhattacharyya:2025gvd}
A.~Bhattacharyya, S.~Ghosh, S.~Pal, and A.~Vinod, ``{Wormholes in finite cutoff JT gravity: A study of baby universes and (Krylov) complexity},'' \href{https://arxiv.org/abs/2502.13208}{{\ttfamily arXiv:2502.13208 [hep-th]}}.

\bibitem{Skenderis:2009ju}
K.~Skenderis and B.~C. van Rees, ``{Holography and wormholes in 2+1 dimensions},'' \href{https://dx.doi.org/10.1007/s00220-010-1163-z}{{\em Commun. Math. Phys.} {\bfseries 301} (2011) 583--626}, \href{https://arxiv.org/abs/0912.2090}{{\ttfamily arXiv:0912.2090 [hep-th]}}.

\bibitem{Zolfi:2023bdp}
H.~Zolfi, ``{Complexity and Multi-boundary Wormholes in 2 + 1 dimensions},'' \href{https://dx.doi.org/10.1007/JHEP04(2023)076}{{\em JHEP} {\bfseries 04} (2023) 076}, \href{https://arxiv.org/abs/2302.07522}{{\ttfamily arXiv:2302.07522 [hep-th]}}.

\bibitem{mir}
M.~Mirzakhani, ``{Weil-Petersson volumes and intersection theory on the moduli space of curves},'' \href{https://dx.doi.org/10.1090/S0894-0347-06-00526-1}{{\em J. Am. Math. Soc.} {\bfseries 20} no.~01, (2007) 1--24}.

\bibitem{mir1}
M.~Mirzakhani, ``{Simple geodesics and Weil-Petersson volumes of moduli spaces of bordered Riemann surfaces},'' \href{https://dx.doi.org/10.1007/s00222-006-0013-2}{{\em Invent. Math.} {\bfseries 167} no.~1, (2006) 179--222}.

\bibitem{kon}
M.~Kontsevich, ``Intersection theory on the moduli space of curves and the matrix airy function,'' {\em Communications in Mathematical Physics} {\bfseries 147} (1992) 1--23.

\bibitem{do}
N.~N.~V. Do, ``Intersection theory on moduli space of curves via hyperbolic geometry,'' {\em PhD Thesis, The University of Melbourne,} (2008) .

\bibitem{wc}
J.~H. Whitehead, ``On equivalent sets of elements in a free group,'' {\em Annals of mathematics} {\bfseries 37} no.~4, (1936) 782--800.

\bibitem{penner}
R.~C. Penner, ``Perturbative series and the moduli space of riemann surfaces,'' {\em Journal of Differential Geometry} {\bfseries 27} no.~1, (1988) 35--53.

\bibitem{Saad:2022kfe}
P.~Saad, D.~Stanford, Z.~Yang, and S.~Yao, ``{A convergent genus expansion for the plateau},'' \href{https://arxiv.org/abs/2210.11565}{{\ttfamily arXiv:2210.11565 [hep-th]}}.

\end{thebibliography}\endgroup

\end{document}